\documentclass[%
 reprint, 
superscriptaddress,
 amsmath,amssymb,
 aps, physrev,
floatfix,
]{revtex4-2}
\usepackage{xcolor}
\usepackage{braket}
\usepackage[nice]{nicefrac}
\usepackage{mathrsfs}
\usepackage{graphicx}
\usepackage{dcolumn}
\usepackage{bm}
\usepackage{fancyhdr}
\usepackage{hyperref}

\begin{document}


\title{\textbf{Noise-Aware Entanglement Generation Protocols for Superconducting Qubits with impedance-matched FBAR Transducers} 
}%

\author{Erin Sheridan}
\thanks{Corresponding author: erin.sheridan.1@afrl.af.mil}
 \affiliation{ U.S. Air Force Research Laboratory, Information Directorate, Rome, NY 13441}

 \author{Michael Senatore}
 \affiliation{ U.S. Air Force Research Laboratory, Information Directorate, Rome, NY 13441}

\author{Samuel Schwab}
 \affiliation{ U.S. Air Force Research Laboratory, Information Directorate, Rome, NY 13441}
 
\author{Eric Aspling}
 \affiliation{Griffiss Institute, Rome, NY 13441}

\author{Taylor Wagner}
 \affiliation{Technergetics, Rome, NY 13441}

\author{James Schneeloch}
 \affiliation{ U.S. Air Force Research Laboratory, Information Directorate, Rome, NY 13441}

\author{Stephen McCoy}
 \affiliation{Booz Allen Hamilton, Rome, NY 13441}

\author{Daniel Campbell}
 \affiliation{ U.S. Air Force Research Laboratory, Information Directorate, Rome, NY 13441}

 \author{David Hucul}
 \affiliation{ U.S. Air Force Research Laboratory, Information Directorate, Rome, NY 13441}
 
 \author{Zachary Smith}
 \affiliation{ U.S. Air Force Research Laboratory, Information Directorate, Rome, NY 13441}

  \author{Matthew LaHaye}
 \affiliation{ U.S. Air Force Research Laboratory, Information Directorate, Rome, NY 13441}

\date{\today}

\begin{abstract}
Connecting superconducting quantum processors to telecommunications-wavelength quantum networks is critically necessary to enable distributed quantum computing, secure communications, and other applications. Optically-mediated entanglement heralding protocols offer a near-term solution that can succeed with imperfect components, including sub-unity efficiency microwave-optical quantum transducers. The viability and performance of these protocols relies heavily on the properties of the transducers used: the conversion efficiency, resonator lifetimes, and added noise in the transducer directly influence the achievable entanglement generation rate and fidelity of an entanglement generation protocol. Here, we use an extended Butterworth-van Dyke (BVD) model to optimize the conversion efficiency and added noise of a Thin Film Bulk Acoustic Resonator (FBAR) piezo-optomechanical transducer. We use the outputs from this model to calculate the fidelity of one-photon and two-photon entanglement heralding protocols in a variety of operating regimes. For transducers with matching circuits designed to either minimize the added noise or maximize conversion efficiency, we theoretically estimate that entanglement generation rates of greater than $160\;\mathrm{kHz}$ can be achieved at moderate pump powers with fidelities of $>90\%$. This is the first time a BVD equivalent circuit model is used to both optimize the performance of an FBAR transducer and to directly inform the design and implementation of an entanglement generation protocol. These results can be applied in the near term to realize quantum networks of superconducting qubits with realistic experimental parameters.

\end{abstract}

\maketitle
\thispagestyle{fancy}

\section{\label{Intro}Introduction}

Microwave-optical quantum transducers \cite{Lambert2020, Lauk2020, Han2021} are critical components for connecting superconducting qubits to quantum networks \cite{Wehner2018, Guccione2020, Ang2024}, toward distributed quantum computing \cite{Cacciapuoti2020}, secure communications \cite{Larson2000}, and other applications requiring entanglement distribution between heterogeneous network nodes. These devices are tasked with coherently converting quantum information between frequencies separated by five orders of magnitude, from $\sim5 \; \mathrm{GHz}$ to $\sim193 \; \mathrm{THz}$. 
In contrast to deterministic entanglement distribution protocols, optically heralded schemes \cite{Duan2001, Barrett2005} can operate effectively with low-efficiency transducers \cite{Wolf2007, Krastanov2021}. While heralded entanglement demonstrations have been carried out with atom-based qubits \cite{Moehring2007, Maunz2009, Olmschenk2009, Hucul2015, Krutyanskiy2023} and optically active solid-state qubits \cite{Hensen2015, Pompili2021, Stolk2024}, where flying optical photons entangled to each remote node mediate the generation of entanglement between them, the same cannot be said for superconducting qubits. In recent years, microwave-optical transducers have improved enough, in principle, to allow initial demonstrations of heralded photon entanglement between superconducting quantum systems via a telecom fiber link. Though entanglement between remote superconducting qubits has been established using microwave interconnects \cite{Narla2016, Kurpiers2018, Dickel2018}, there has been no experimental demonstration of optically-mediated entanglement generation between remote superconducting quantum circuits to date. This may be because previous efforts to integrate superconducting qubits with microwave-optical transducers have been limited by prohibitively low qubit coherence times and high added noise due to heat dissipation in the transducer \cite{Mirhosseini2020, vanThiel2023}, as well as limited cooling power in cryogenic systems. 

\begin{figure*}
\includegraphics[width =\textwidth] {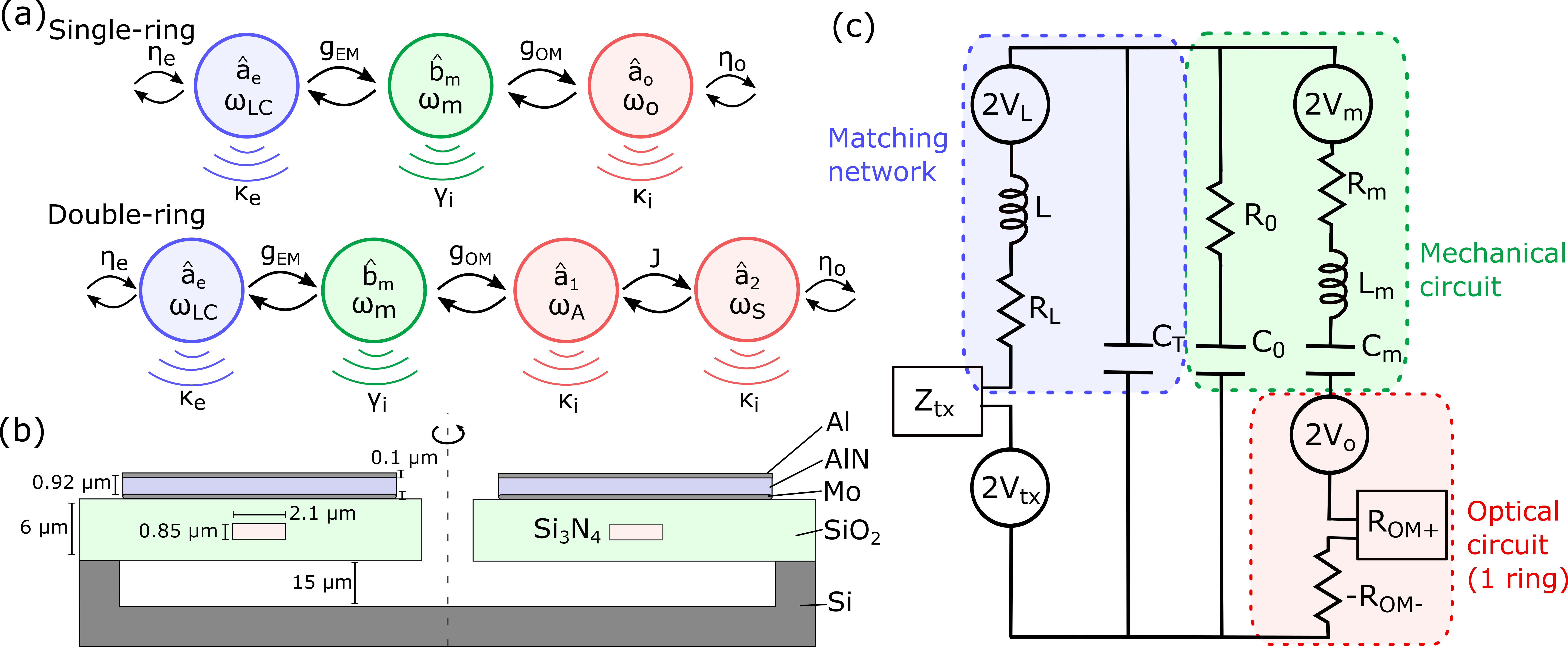}
\caption{ (a) Schematic of the constituent modes of a piezo-optomechanical transducer and their interactions with each other and the outside world for (top) a single-ring device and (bottom) a double-ring device. $g_{EM}$ is the electromechanical coupling rate, $g_{OM}$ is the optomechanical coupling rate, $\eta_e$ is the microwave signal coupling efficiency, $\kappa_e$ is the microwave mode ($\hat{a}_e$) loss  $J$ is the optical-optical coupling rate, rate, $\gamma_i$ is the mechanical mode ($\hat{b}_m$) intrinsic linewidth, $\kappa_i$ is the optical mode ($\hat{a}_o$, $\hat{a}_A$, $\hat{a}_S$ ) intrinsic linewidth, and $\eta_{o}$ is the optical mode external coupling efficiency. (b) Side profile of an example material stack for an FBAR transducer, see\cite{Blesin2021}. The dashed line denotes the axis of rotational symmetry. General estimates of layer thicknesses are shown, and should be understood as rough guides. Specific FBAR devices may have slightly modified dimensions. (c) Th\'{e}venin equivalent circuit for a piezo-optomechanical transducer, including a matching network.  $Z_{tx}$ is the input impedance, $L$ is the matching inductor, $C_T$ is the matching capacitor, $R_L$ is the matching inductor's resistance, $C_0$ is the static capacitance, $R_0$ is the static resistance, $R_m, L_m, C_m$ are the motional resistance, inductance, and capacitance, respectively, $R_{OM\pm}$ are the equivalent resistances for the optical signal and noise outputs, and $V_{tx}, V_{L}, V_{m}, V_{o}$  are the Th\'{e}venin equivalent voltage sources which accompany every resistive element. See \cite{Wu2020}. }
\label{fig:Fig1}
\end{figure*}

Piezo-optomechanical transducers are reaching a point in their maturity where experimental demonstrations of  heralded entanglement protocols \cite{Zeuthen2020, Zhong2020, Zhong2020a, dAvossa, Ang2024} are possible.   As shown in Figure \ref{fig:Fig1}(a), these transducers achieve coherent microwave-optical conversion by first converting a microwave photon to an acoustic phonon via the piezoelectric effect, with electromechanical coupling rate $g_{EM}$, then converting the phonon to an optical photon via radiation pressure or the stress-optical effect, with a cavity-enhanced optomechanical coupling rate $g_{OM}$. In the next section, we will introduce a specific class of piezo-optomechanical transducers. We will then proceed to design impedance-matching circuits for these devices, then explore their operation within entanglement heralding protocols.

\subsection{\label{Bhave Intro}The FBAR Piezo-optomechanical transducer}
Transducers based on optomechanical crystal  (OMC) cavities have achieved high conversion efficiencies\cite{vanThiel2023, Meesala2024, Weaver2024}. However, these devices suffer from high added noise quanta per transduced photon, as the OMC cavity is susceptible to optical absorption-based heating \cite{Ren2020}, leading to non-negligible thermal Brownian motion of the mechanical modes \cite{Mirhosseini2020}. Recently, two-dimensional OMCs have been developed to enhance thermalization \cite{Mayor2024,sonar2024} and have achieved thermal occupancies of the phonon cavity $<1$ at high optical pump repetition rates with optical cavity occupations on the order of $10^3$ photons. These results motivate the exploration of bulk acoustic resonators for even better thermalization, promising acoustic cavity ground state operation at even higher optical cavity occupations.

As an alternative to transducers based on OMCs, transducers utilizing bulk acoustic resonators, such as the Thin Film Bulk Acoustic Resonator (FBAR) devices \cite{Tian2020,Blesin2021,blesin2023,Schneeloch2025} promise superior power handling capabilities \cite{Liu2020,Vinita2024, Doeleman2023} and high electromechanical extraction and phonon injection efficiencies \cite{tian2024}, in addition to the advantage of their compatibility with CMOS foundry fabrication. The bulk nature of these devices may allow for higher pump powers and repetition rates while maintaining fewer added noise quanta per transduced photon, since heat can propagate through the bulk of the chip. Bulk acoustic modes have been integrated with superconducting circuits in the past and have achieved strong coherent coupling \cite{Chu2017, Chu2018}.  Furthermore, the degeneracy of the microwave and mechanical modes in FBAR devices eliminates the need to tune the microwave and mechanical modes into resonance, using magnetic fields from e.g. microwave coils installed within a dilution refrigerator \cite{Weaver2024}, significantly reducing the complexity of operation. 

In this work, we design impedance-matching networks and entanglement heralding protocols for the FBAR transducer, following the work done by Wu \emph{et al.} \cite{Wu2020} on matching network design and by Zeuthen \emph{et al.}  \cite{Zeuthen2020} on heralding protocol design. A side profile of an FBAR transducer is shown in Figure \ref{fig:Fig1}(b). These converters utilize high-overtone bulk acoustic resonances (HBARs) as an intermediary between microwave and optical modes. For microwave-to-optical conversion, microwave signals encounter a microwave cavity consisting of an aluminum nitride (AlN) piezoelectric film sandwiched between signal and ground electrodes. During the electromechanical interaction, these electrical signals generate expansion and contraction of the film through the converse piezoelectric effect; these expansions and contractions in turn launch acoustic waves into the transducer’s acoustic cavity. When the cavity length is equal to an integer multiple of the acoustic wavelength, multiple acoustic standing  waves interfere constructively, leading to a resonant feature (an HBAR mode). The optomechanical interaction arises from the acoustic resonances modifying the index of refraction of an optical micro-ring resonator (MRR) via the stress-optical effect.  


Intermediate-mode transducers, including piezo-optomechanical devices, can be described by the generic interaction Hamiltonian
\begin{multline}
H_{int} = \\ \hbar \sum_{i=e,o} g_{iM}(\hat{b}_m e^{-i\omega_m t}+\hat{b}^{\dag}_m e^{i\omega_m t})(\hat{a}_i e^{-i\Delta_i t}+\hat{a}^{\dag}_i e^{i\Delta_i t})\\
\end{multline}
for the mechanical mode $\hat{b}_m$ with angular frequency $\omega_m$, coupling rates $g_{iM}$ and a pump-cavity detuning defined as $\Delta_i = \omega_{i}-\omega_{pump}$ for modes $\hat{a}_i$.  In FBAR transducers the mechanical mode and microwave mode are degenerate, so the mechanical field can be taken as pumped directly by the microwave field  with an input-output relation of $\hat{a}_{e,in}+ \hat{a}_{e,out} = \sqrt{\gamma_{ex}} \hat{b}_m$ \cite{Blesin2021, Schneeloch2023} for a microwave photon-phonon coupling rate $\gamma_{ex}\sim\nicefrac{4g_{EM}^2}{\Gamma}$ with microwave cavity linewidth $\Gamma$. This simplifies the interaction Hamiltonian, which can now be written as
\begin{multline}
H_{int} = \hbar g_{OM}(\hat{b}_m e^{-i\omega_m t}+\hat{b}^{\dag}_m e^{i\omega_m t})(\hat{a}_o e^{-i\Delta_o t}+\hat{a}^{\dag}_o e^{i\Delta_o t})
\end{multline}
with pump-optical mode detuning $\Delta_o$. We can select one of two possible optomechanical interactions by red-detuning our pump laser pulse such that $\Delta_o = \omega_m$ or blue-detuning the pulse such that $\Delta_o = -\omega_m$ \cite{Lauk2020}. Multiplying out the optomechanical interaction terms and discarding high-frequency terms oscillating at $\pm 2 \omega_m$, we have, for the red-detuned case,
\begin{multline}
\cr H_{int,red} =  \hbar g_{OM}(\hat{b}_m^{\dag}\hat{a}_o+\hat{b}_m\hat{a}_o^{\dag}) \cr
\label{eq:hint_red}
\end{multline}
yielding a beamsplitter-like interaction between the optical cavity mode and the mechanical mode, where excitations are swapped between modes. The red-detuned optical pump, together with a resonant microwave pump, enables direct microwave-to-optical transduction. A blue-detuned pump $\Delta_o = \omega_m$ leads to an optomechanical two-mode squeezing interaction:
\begin{multline}
\cr H_{int,blue} =  \hbar g_{OM}(\hat{b}_m\hat{a}_o+\hat{b}_m^{\dag}\hat{a}_o^{\dag}) \cr
\label{eq:hint_blue}
\end{multline}
which will not be the focus of this work. Finally, if we incorporate a second optical mode, we re-label the modes as $\hat{a}_{1}$ and $\hat{a}_2$ coupled with rate $J$. We can write the complete red-detuned interaction Hamiltonian as \cite{Blesin2021}
\begin{multline}
\cr H_{int,red} = \hbar g_{OM}(\hat{b}_m^{\dag}\hat{a}_1+\hat{b}_m\hat{a}_1^{\dag})+ \hbar J(\hat{a}_1^{\dag}\hat{a}_{2}+\hat{a}_1\hat{a}_{2}^{\dag}).\cr
\end{multline}

More details on the theoretical foundations underpinning FBAR transducer operation can be found in references \cite{Tian2020,Blesin2021,blesin2023,Schneeloch2023}.

\subsection{\label{IM Intro}Butterworth-van Dyke model and impedance matching}
A disadvantage of FBAR transducers is their comparatively low optomechanical coupling rates  \cite{tian2024}. An estimate based on finite-element simulations for a layout similar to that shown in Figure \ref{fig:Fig1}(b) puts the single-photon optomechanical coupling rate at $400 \; \mathrm{Hz}$  \cite{Blesin2021}. This low coupling rate limits the conversion efficiency of the device for low optical pump powers. In Blesin \emph{et al.}\cite{Blesin2021}, a peak on-chip microwave-optical conversion efficiency of $40\%$ was theoretically estimated for an optical pump power of $100 \; \mathrm{mW}$. While promising, achieving such high pump powers on chip is unrealistic in a milli-Kelvin cryogenic environment, and would likely introduce large amounts of thermal noise in the transducer that would preclude the ability to coherently transduce quantum information. 

The performance of the FBAR devices can be improved by the use of  electrical impedance-matching networks, which are used to boost signal transfer efficiency \cite{Zeuthen2018,Wu2020}. Impedance matching networks can be used to either maximize the conversion efficiency of the transducer by balancing its cooperativities, or minimize added noise by maximizing the electromechanical coupling rate. To design such networks, we first need to formulate an equivalent circuit representation of the transducer.

The BVD model is an equivalent circuit model that has been used extensively to understand the behavior of FBAR acoustic devices \cite{Larson2000, Uzunov2017, Tian2020, Wu2020, tian2024}. Impedance matching networks can be designed within an expanded  BVD equivalent circuit model that captures the behavior of the entire transducer \cite{Zeuthen2018,Wu2020}. What's more, the physical parameters needed as inputs for the extended BVD model, including the transducer's static capacitance and its electromechanical coupling factor ($k_{eff}^2 = \nicefrac{4 g_{EM^2}}{\omega_m^2}$) for resonant microwave and mechanical modes), can be extracted in a straightforward way from finite-element simulations \cite{Uzunov2017}. In Sec.~\ref{Matching} we use the extended BVD model to design impedance matching networks for FBAR transducers with a single optical cavity and with two coupled optical cavities. 

\begin{figure*}
\includegraphics[width =\textwidth]{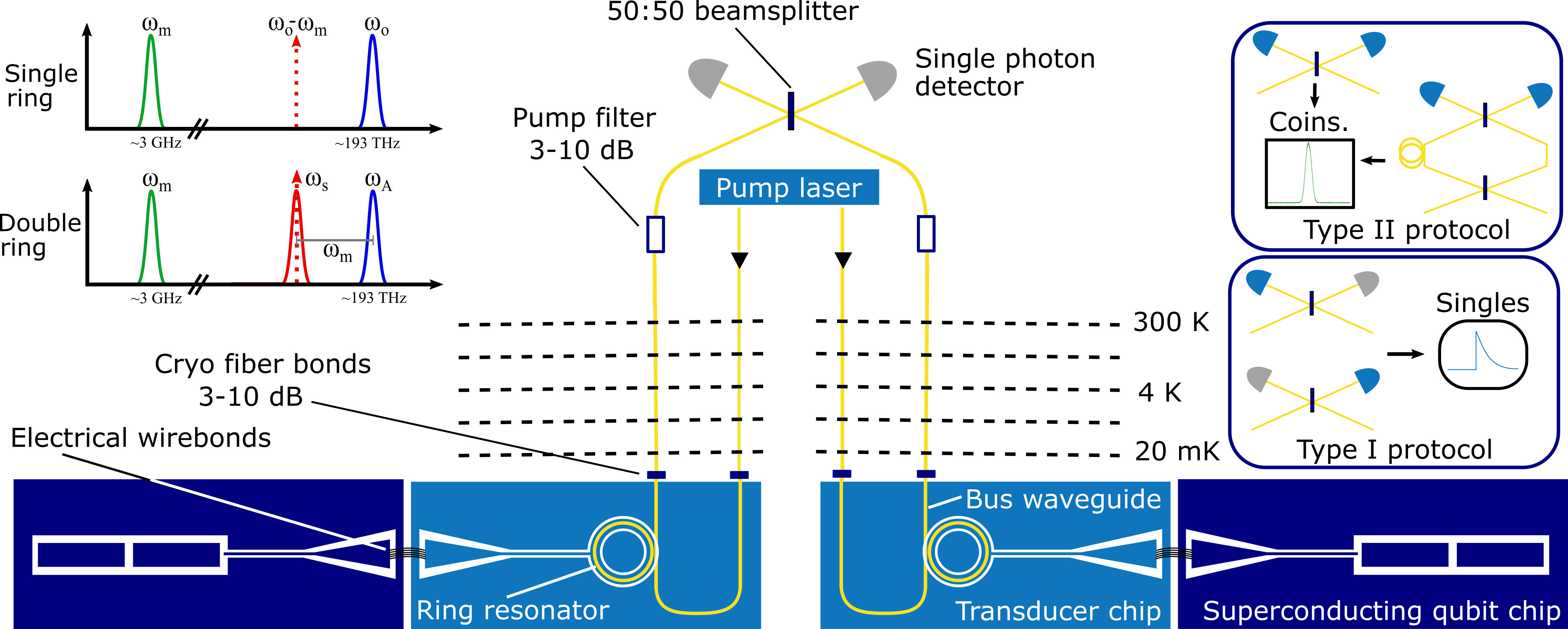}
\caption{Diagram of an entanglement heralding setup between two nominally identical quantum network nodes with superconducting circuits and microwave-optical transducers, which are located within cryostats. A single-ring transducer is depicted. A laser sends red-detuned pump pulses to the transducers, and qubits are initialized to their excited state. Photons emitted by each transducer coupled off-chip via cryogenic fiber bonds, then undergo pump filtering before impinging on a 50:50 beamsplitter. Typical loss values are labeled. Top left: density of states diagrams for red-detuned pumping of single-ring and double-ring FBAR transducers Top right: for the Type I protocol, single photon clicks in either detector herald entanglement between the remote superconducting qubits. For the Type II protocol, coincident detection events (coins.) do the same, using a balanced or unbalanced interferometer.}
\label{fig:Fig2}
\end{figure*}
Here, for the first time, we show that the BVD model can be used not only to optimize an optomechanical transducer, but also to optimize optically-mediated entanglement heralding protocols making use of these transducers. 
\subsection{\label{Heralding Intro}Entanglement heralding for superconducting qubits}
Often referred to as the DLCZ \cite{Duan2001} or Barrett-Kok \cite{Barrett2005} protocol, optically-mediated entanglement of remote matter-based qubits involves interfering flying photons emitted by those qubits in a Bell state analyzer. The beamsplitter within the analyzer erases which-path information before the photon(s) are incident upon single photon detectors. This information erasure allows one to project entanglement onto the remote quantum memories. These protocols can broadly be classified as one-photon (Type I) or two-photon (Type II) protocols \cite{Luo2009}. 

A Type-I protocol is that which uses an optical single-photon state to establish a distributed Bell state between matter-based qubits at remote network nodes. It will only succeed when one photon is emitted, in total, from both network nodes. As detailed below, Type-I protocols are susceptible to sources of infidelity from two-photon emission events. They are also sensitive to path-length variations within the experimental setup, where even slight changes in the length of the network links can change the relative phase of the entangled state, leading to decoherence. In contrast, Type-II protocols use two optical photons, one emitted from each of two nodes, and measure coincident detection events between them.

The phase stability requirement for Type-I protocols \cite{Stockill2017, Pompili2021} incentivizes the development of Type-II protocols, which are more robust to instabilities due to path-length variations. They are also immune to some sources of infidelity that affect Type-I protocols. The cost of these improvements is increased experimental complexity: quantum information is no longer encoded in a number state, but must be encoded in the time-bin, polarization, frequency, or other degrees of freedom of the optical photons\cite{Luo2009}.

When we consider heralding protocols in the context of superconducting qubits and transducers, even more complexities are introduced. As discussed above, the transducer can operate in different regimes depending on the frequency of the optical pump with respect to the optical cavity mode(s). Spontaneous parametric down-conversion (SPDC) operation is induced by an optical pump blue-detuned from the optical cavity (Equation \ref{eq:hint_blue} ), whereas a red-detuned optical pump yields a beam-splitter-type interaction that can be utilized for quantum coherent state transfer (direct transduction, Equation \ref{eq:hint_red}) \cite{Aspelmeyer2014}. SPDC operation generates entangled microwave-optical photon pairs within the transducer, where the microwave photon can then be swapped into a superconducting qubit. Zhong \emph{et al.}\cite{Zhong2020} proposed a Type-II heralding protocol utilizing blue-detuned pumping for SPDC operation of the transducer and time-bin encoded optical photons. Recently, non-classically correlated microwave-telecom photon pair generation has been experimentally demonstrated \cite{Jiang2023, Meesala2024}.

In contrast to SPDC operation, direct operation requires the transducer to capture a microwave photon emitted by a qubit. Direct operation allows us to avoid populating the transducer's acoustic mode with multi-phonon states, which is a significant source of infidelity. In this work we focus on red-detuned transducer operation, as it is experimentally more straightforward to implement and has been shown to generate higher fidelity entanglement generation\cite{Krastanov2021} than SPDC-based heralding protocols for superconducting qubits.  

Zeuthen \emph{et al.}  \cite{Zeuthen2020} previously explored the dependence of entanglement fidelity on the conversion efficiency and added noise associated with a microwave-optical quantum transducer \cite{Zeuthen2020}. Here we build off of that work, using outputs from the extended BVD model to obtain estimates for the fidelity of these protocols for an FBAR transducer in the presence of thermal noise. As a result, for the first time, we show how the BVD model can be used to optimize the transducer's figures of merit and those of a heralding protocol. Insights gleaned from our simulations can be used to precisely engineer transducers for optimal operation within a specific environment and protocol.

While previous circuit-based treatments of FBAR transducers focus on device characterization\cite{Blesin2021, Schneeloch2023, Wu2020}, our approach is the first to directly connect BVD circuit parameters to quantum entanglement protocol performance metrics. This connection allows us to concurrently optimize both conversion efficiency and added noise for entanglement applications, whereas previous work optimized these figures of merit separately.

Our treatment also generalizes the extended BVD model in several key ways: (a) including two optical resonators to model photonic molecule architectures, whose optical emission frequency can be tuned via the symmetric/antisymmetric supermode splitting, (b) including parasitic circuit elements and their compensation in matching network design, and (c) connecting device-level optimization directly to protocol-level performance. Unlike previous theoretical explorations of entanglement heralding protocols \cite{Luo2009, Krastanov2021}, we provide a complete device engineering framework that bridges transducer physics to quantum networking applications.

\section{\label{Matching}Impedance matching networks for FBAR transducers}
Impedance matching is a well-known technique for maximizing power transfer, minimizing signal reflections, or enhancing signal-to-noise in a network of elements with varying impedance.  In the domain of radio frequency and microwave quantum circuitry, this technique has been utilized for an array of applications involving ultra-sensitive measurement of solid-state quantum electronic and electromechanical devices \cite{LaHaye2009, Ares2016}. A common approach in these applications is to utilize impedance matching L-networks, which consist of lumped element inductors and capacitors to cancel the reactive (imaginary) part of the load impedance, while transforming the real part to match the source impedance \cite{Pozar}. Such networks can be used to maximize signal transfer in an optomechanical transducer by forming a bridge between a $50 \; \mathrm{Ohm}$ input impedance and a higher-impedance piezoelectric element. 

When we apply this idea to optomechanical transducers, we need to think somewhat differently about the quantities we need to match. Specifically, we want the matching network to match the electromechanical loading and the optomechanical loading on the acoustic mode. At the same time, the resonant frequency of the matching circuit must match that of the acoustic mode \cite{Wu2020}. To determine the parameters necessary to achieve these conditions simultaneously, we utilize an equivalent circuit representation constructed from the BVD model, as described below.
\begin{figure}
\includegraphics{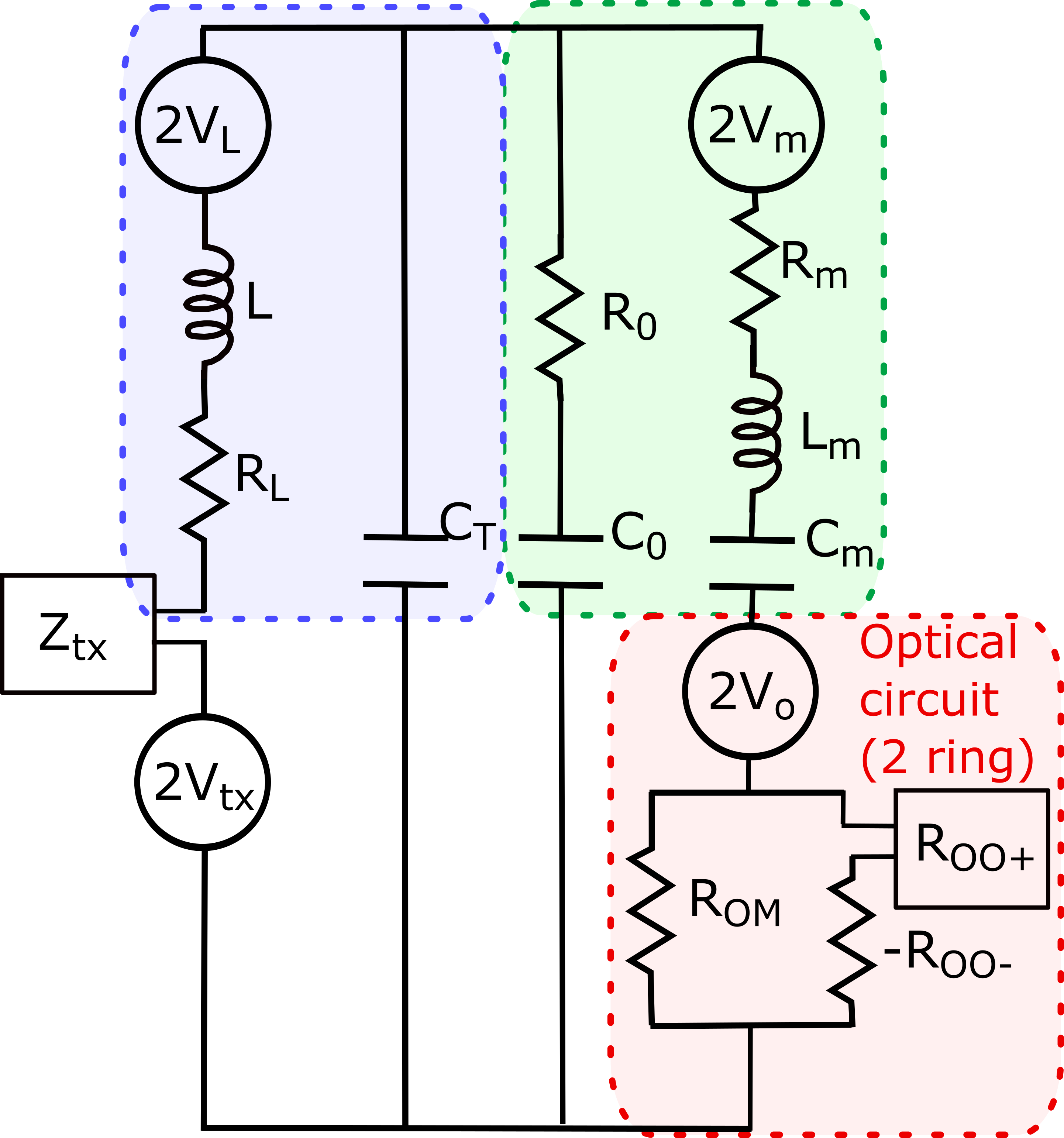}
\caption{A modified equivalent circuit for a transducer with two optical ring resonators. The three sub-circuits are the matching network (blue), mechanical sub-circuit (green) and the optical sub-circuit (red). As compared to the equivalent circuit shown in Figure 1(c), the optical sub-circuit has been changed here. The optomechanical coupling is now represented by the effective resistance $R_{OM}$ and the effective resistors $R_{OO \pm}$, representing the upper and lower optical sidebands, have been added in parallel with $R_{OM}$.}
\label{fig:Fig3}
\end{figure}

\subsection{\label{BVD details}The BVD model}
A BVD equivalent circuit for a piezo-mechanical element consists of both a static arm and a motional arm. Within the static arm, the static capacitor $C_0$ captures the usual capacitive behavior of the device's electrodes. The mechanical motion is described by the motional arm, the motional inductance $L_m$ and motional capacitance $C_m$ capture the mechanical kinetic and potential energy, respectively, and a motional resistance $R_m$ represents the dissipation of acoustic waves. 

Generally, within the BVD model, one defines the electromechanical coupling factor as \cite{Chang1995}.
\begin{equation}
    k_{eff}^2 = C_m/(C_0+C_m)
\end{equation}
However, as we will show in Sec.~\ref{matching}, for FBAR transducers $C_m \ll C_0$ and as such, we define the motional elements in the familiar form:
\begin{equation}
    C_m = k_{eff}^2 C_0
\end{equation}
\begin{equation}
    L_m = \frac{1}{\omega_m^2 C_m}
\end{equation}
\begin{equation}
    R_m = \gamma_m L_m
\end{equation}
with acoustic mode $\omega_m$ that exhibits an intrinsic linewidth $\gamma_m$.

The behavior of a piezo-mechanical device is understood by examining its admittance 
\begin{equation}
    Y(\omega) = i \omega C_0 + \frac{1}{L_m}\frac{i \omega}{-\omega^2 +i\omega\gamma_m+\omega_m^2}.
\end{equation}
Near resonance with a mechanical mode at $\omega_m$, the imaginary part of the admittance curve will exhibit both a peak and a dip. The peak appears at the series resonant frequency $\omega_s$ and the dip appears at the parallel resonant frequency $\omega_p$. The electromechanical coupling factor $k_{eff}^2$ can be defined with respect to the series-parallel frequency separation:
\begin{equation}
    k_{eff}^2 = \frac{\omega_p^2-\omega_s^2}{\omega_p^2}.
\end{equation}
One might notice that for our previous condition $C_m \ll C_0$ we find $k_{eff} \ll 1$ and $\omega_s \approx \omega_p$. The resonant frequency relates to the series and parallel frequencies by $\omega_m = (\omega_s+\omega_p)/2$, and therefore a reasonable approximation exists as $\omega_m \approx \omega_s.$ This approximation will prove to be beneficial in the next section.
Far away from resonance, the device behaves as a capacitor, so its capacitance can be extracted from the admittance curve using the relation $Im(Y_{\omega}) \rightarrow \omega C_0$. As such, running a finite-element simulation of a device and plotting its admittance curve allows us to extract $k_{eff}^2$ and $C_0$ for our specific device geometry and inform the equivalent circuit model.
Moving beyond the standard BVD model, the modified BVD (mBVD) model includes an additional static resistance $R_0$ to the static arm of the piezo-mechanical device, which captures the dielectric loss due to energy dissipation in the piezoelectric layer. The mBVD model has been found to more accurately model the behavior of FBAR devices\cite{Larson2000}, so we use it here. See Appendix~\ref{app:staticR} for more details. 

\subsection{\label{extended BVD details}The Extended BVD model for optomechanical transducers}

We can use an extended BVD model to go beyond the piezo-mechanical element and build an equivalent circuit for the entire transducer device, including its optical elements. Using the framework developed by Zeuthen \emph{et al.}\cite{Zeuthen2018} and Wu \emph{et al.}\cite{Wu2020}, we developed an equivalent circuit to model the FBAR transducer. Like the device considered in \cite{Wu2020}, the FBAR transducer utilizes an optomechanical supermode, where a supermode is formed by the hybridization of two modes that are nearly resonant with each other. In the FBAR transducer, the piezoelectric and mechanical modes are resonant \cite{Sridaran2011}, as discussed in Section  \ref{Bhave Intro}. 

The equivalent circuit for a piezo-optomechanical transducer consists of three sub-circuits: the electrical matching network sub-circuit, the piezo-mechanical sub-circuit, and the optical sub-circuit, all of which are coupled together. Th\'{e}venin equivalent circuit diagrams are shown in Figures~\ref{fig:Fig1} and ~\ref{fig:Fig3}. While we refer the interested reader to references \cite{Zeuthen2018, Wu2020} for a detailed derivation of the extended BVD model, we briefly introduce the formalism here. 

Before moving on, it's important to clarify an assumption that is made in this work and in previous work on this model \cite{Zeuthen2018, WangJiang2022} :  we are assuming the electrical and optical modes are strongly coupled to a single acoustic mode, and weakly coupled to all other acoustic modes. In practice this may not be the case, particularly in unreleased FBAR transducers with a dense forest of acoustic modes with a free spectral range of $<20\;\mathrm{MHz}$ \cite{Tian2020}. Generalizing the equivalent electrical circuits for transducers with multiple acoustic modes coupled in parallel is an area of potential future exploration.

\subsubsection{\label{matching}The matching circuit}
A matching network (blue sub-circuit in Figures \ref{fig:Fig1}, \ref{fig:Fig3}) consists of an inductor $L$, with resistance $R_L$, and a capacitor $C_T$ coupled to the transducer's acoustic mode via its static capacitance $C_0$. The matching circuit has a resonant frequency of
\begin{equation}
    \omega_{LC} = \frac{1}{\sqrt{L(C_0+C_T})}
\end{equation}
and a quality factor
\begin{equation}
    Q_{LC} = \frac{1}{Z_{tx}+R_L}\sqrt{\frac{L}{C_0 + C_T}}
\end{equation}
with input impedance $Z_{tx}$ from a transmission line. Its linewidth $\kappa_e$ is
\begin{equation}
    \kappa_e = \frac{Z_{tx}+R_L}{L}
\end{equation}
and its external coupling efficiency $\eta_e$ is
\begin{equation}
    \eta_e = \frac{Z_{tx}}{Z_{tx}+R_L}.
\end{equation}
For a superconducting circuit, we can set $R_L=0$. This is a fairly accurate assumption that is explored in detail in Appendix~\ref{app:losss}.

Now that we've defined the matching circuit, we can write an expression for the electromechanical loading of the acoustic mode due to its presence:
\begin{equation}
    R_{EM} = Q_{LC}^2(Z_{tx}+R_L).
\end{equation}
This loading leads to an electromechanical cooperativity of
\begin{equation}
    \mathcal{C}_{EM} = \frac{R_{EM}}{R_m} = \frac{4g_{EM}^2}{\gamma_m \kappa_e} 
\end{equation}
where the electromechanical coupling rate is 
\begin{equation}
    g_{EM} = \frac{\sqrt{k_T^2}\omega_m}{2} \; \; :\; \;k_T^2 = \frac{C_m}{C_m+C_0+C_T}.
\end{equation}
The electromechanical cooperativity captures the strength of coherent coupling between the electrical and mechanical modes. We can understand the action of the matching network as an impedance transformation provided by the resonant signal enhancement due to its loaded quality factor $Q_{LC}$. The presence of the electrical LC circuit modifies the resonance condition for the acoustic mode:
\begin{equation}\label{eq:resonant omega}
    \omega_{m} = \sqrt{\frac{1}{L_m}\left( \frac{1}{C_m}+\frac{1}{C_T+C_0}\right)}
\end{equation}
and, as elsewhere, the presence of the optical mode is assumed not to shift the mechanical resonance \cite{Wu2020}.

\subsubsection{The Optomechanical sub-circuit}
Next we consider the optomechanical cavity, which is coupled in series to the motional part of the piezo-mechanical element. The optical MRR is understood to have a total linewidth $\kappa_o = \kappa_i + \kappa_{ext}$, where $\kappa_i$ is its intrinsic linewidth and $\kappa_{ext}$ is its external coupling rate. The optical external coupling efficiency is given by
\begin{equation}
\eta_o = \frac{\kappa_{ext}}{\kappa_o}.
\end{equation}
The optomechanical coupling between this optical mode and the mechanical mode is represented by frequency-independent effective resistances $R_{OM,\pm}$. The positive resistance $R_{OM,+}$ represents the desired output mode (upper sideband for direct transduction), and the negative resistance $-R_{OM,-}$ represents the optical noise channel due to the Stokes process creating photons in the lower sideband, which excite spurious phonons in the mechanical mode, creating noise photons in the upper sideband \cite{Wu2020}.
\begin{multline}
\cr
    R_{OM, +} = R_m \mathcal{C}_{OM} \mathscr{L}^2_+ \\
    R_{OM, -} = R_m  \mathcal{C}_{OM} \mathscr{L}^2_- \cr
\end{multline}
Here $ \mathcal{C}_{OM}= \frac{4g_{OM}^2}{\gamma_m \kappa_o}$ is the optomechanical cooperativity and the Lorentzian sideband amplitudes $\mathscr{L}^2_{\pm}$ are
\begin{equation}
    \mathscr{L}^2_{\pm} = \frac{(\kappa_o/2)^2}{(\kappa_o/2)^2+(\omega_m \pm \Delta)^2}
\end{equation}
where $\Delta$ is the detuning of the optical pump from the optical cavity frequency. The optomechanical loading on the acoustic mode is then
\begin{equation}
    R_{EM}^{opt} = R_m+R_{OM,+}-R_{OM,-}.
\end{equation}
Now that we have defined expressions for each sub-circuit of the transducer, let's put the sub-circuits together to characterize the performance of the overall device.

\subsubsection{\label{figs of merit}Figures of Merit}
Given the formalism established above, we can write an expression for the conversion efficiency of the transducer. The conversion efficiency can be written heuristically as
\begin{equation}
    \eta = \eta_{e} \eta_{o} \eta_{int}
\end{equation}
where $\eta_{e,o}$ are the external coupling efficiencies of the electrical and optical cavities, respectively, and $\eta_{int}$ is defined as the internal efficiency of the device. The form of the expression for $\eta_{int}$ depends on the number of coupled modes comprising the transducer \cite{Han2021}. For a standard optomechanical device with cooperativites $\mathcal{C}_{ij}$, $\mathcal{C}_{jk}$ between modes $i,j,k$, we write 
\begin{equation}
\eta_{int}= \frac{4\mathcal{C}_{ij}\mathcal{C}_{jk}}{(1+\mathcal{C}_{ij}+\mathcal{C}_{jk})^2}.
\end{equation} 
From our equivalent circuit we can write the appropriate expression for our device:
\begin{multline}
    \label{eta_1ring}
   \eta = \eta_e \eta_o \frac{4R_{EM}R_{OM,+}}{(R_m+R_{EM}+R_{OM,+}-R_{OM,-})^2}\\ 
   = \eta_e \eta_o \frac{4 \mathcal{C}_{EM} \mathcal{C}_{OM}\mathscr{L}^2_+}{(1+ \mathcal{C}_{EM}+ \mathcal{C}_{OM}(\mathscr{L}^2_+-\mathscr{L}^2_-))^2}.
\end{multline}
While Equation~\ref{eta_1ring} is the commonly cited, or standard, transduction efficiency expression for conversion with one intermediate mode, it only pertains in the low coupling limit, where $C_{EM}\;,C_{OM} < 1$ \cite{WangJiang2022}. We must take note of this, since the presence of the matching circuit will result in large electromechanical cooperativity $C_{EM} \gg 1$. As such, we must also consider an alternative efficiency expression \cite{WangJiang2022}:
\begin{multline}
    \label{eta_mod_1ring}
   \eta_{alt}  = \eta_e \eta_o \frac{R_{OM+}}{R_{OM,+}-R_{OM,-}+R_m}\\
   =\eta_e \eta_o \frac{ \mathcal{C}_{OM}\mathscr{L}^2_+}
   {\left( \mathcal{C}_{OM}(\mathscr{L}^2_+-\mathscr{L}^2_-)+1\right)}
\end{multline}

Equation~\ref{eta_mod_1ring} reflects the fact that $ \mathcal{C}_{OM}$ is much smaller than $ \mathcal{C}_{EM}$ for our device, and as the bottleneck, it determines the conversion efficiency. In Table \ref{tab:results} below, we see that for $ \mathcal{C}_{EM}>1$, the standard expression for the efficiency disagrees with the alternative expression. A downside to the alternative expression is its independence from the matching network: according to it, the conversion efficiency does not see a boost from the impedance matching process. As such, we include efficiency estimates from both the standard and alternative expressions in Table~\ref{tab:results}.

From the equivalent circuit we can also extract expressions for added noise during transduction, due to the presence of both optical amplification noise and thermo-mechanical noise. Optical amplification noise, or Raman noise, occurs when a pump photon  produces an output photon at the angular frequency $\omega_{pump}-\omega_m$ via a Stokes process. From two-mode squeezing, this process also produces a phonon which is then transduced into the upper sideband as an optical Raman noise photon.
\begin{equation} \label{no}
    n_o = \frac{1}{\eta_e}\frac{ \mathcal{C}_{OM}\mathscr{L}^2_-}{ \mathcal{C}_{EM}}
\end{equation}

The number of thermo-mechanical noise quanta is given by 
\begin{equation}\label{nth}
    n_{th} = \frac{1}{\eta_e} \frac{n_m}{ \mathcal{C}_{EM}}
 \end{equation}
Here  $n_m$ is the thermal bath occupancy given by the Bose-Einstein distribution, $n_m = (e^{\hbar \omega / k_B T}-1)^{-1}$. The quantity $\nicefrac{ \mathcal{C}_{EM}}{n_m}$ is the electromechanical quantum cooperativity. It can be understood as the ratio of coherent electromechanical coupling to the thermal decoherence induced by the mechanical bath, and it reveals how the impedance matching circuit enhances the ratio of electrical signal to mechanical thermal noise in the transducer.

Finally, the conversion bandwidth is given by the dynamically broadened mechanical linewidth:
\begin{multline}
  \Delta \omega = \gamma_m (1+  \mathcal{C}_{EM} +  \mathcal{C}_{OM}(\mathscr{L}^2_+-\mathscr{L}^2_-)) \\ = (R_m+R_{EM}+R_{OM,_+}-R_{OM,-})/L_m.
\end{multline}

Details of how we modified the extended BVD model to include the static resistance $R_0$ can be found in Appendix \ref{app:staticR}. Now that we have defined the figures of merit of our transducer as they relate to our model, we can proceed to design an impedance matching network.  

\subsubsection{\label{design_match}Designing the matching network}
It is ideal for all of the figures of merit for our transducer to be optimized simultaneously. In practice, there will be tradeoffs, and devices will have to be designed with use cases and prioritized properties in mind. 

Let us begin with the most straightforward goal: optimizing the microwave-to-optical conversion efficiency of the transducer. If we assume the optomechanical and mechanical parameters to be fixed, the signal transfer efficiency expression (Equation~\ref{eta_1ring}) reaches its maximum as a function of $C_{EM}$ when
\begin{equation} \label{eq:max_eta_coop}
     \mathcal{C}_{EM} =  \mathcal{C}_{EM}^{opt} = 1 +  \mathcal{C}_{OM}(\mathscr{L}^2_+-\mathscr{L}^2_-).
\end{equation}
To satisfy the cooperativity relationship in Equation~\ref{eq:max_eta_coop} we need to choose the matching circuit parameters $C_T$, $L$, such that the electromechanical loading $R_{EM}$ is equal to the optomechanical loading $R_{EM}^{opt}$\cite{Wu2020}:
\begin{equation}
    R_{EM} = R_m+R_{OM,+}-R_{OM,-}.
\end{equation}
 This amounts to choosing the electromechanical broadening of the mechanical mode to be equal to the intrinsic mechanical linewidth plus the net optomechanical broadening. Working through the algebra, this relationship leads us to two equations for the matching circuit that we must solve in accordance with the resonance condition $\omega_{LC} = \omega_m$:
\begin{equation}\label{eq:CT general}
    C_T = \frac{1}{\omega_m}\sqrt{\frac{1}{R_{EM}^{opt}(Z_{tx}+R_L)}}-C_0
\end{equation}
\begin{equation}\label{eq:L general}
   L = \frac{1}{\omega_m}\sqrt{R_{EM}^{opt}(Z_{tx}+R_L}).
\end{equation}
This system of equations must be solved self-consistently. However, an analytical solution of Equations.~\ref{eq:CT general} and~\ref{eq:L general} may be obtained if we assume that the dependence of $R^{opt}_{EM}$ on the acoustic mode frequency $\omega_m$ can be neglected. This is a valid assumption when $k_{eff}^2 \ll \kappa_o / \omega_s$ \cite{Wu2020}. For the transducer,  $k_{eff}^2 = 4.3 \times 10^{-3}$ and $\kappa_o / \omega_s = 4.566 \times 10^{-2}$, so we can proceed with this approximation.
For the analytical solution at the resonance condition $\omega_{LC} = \omega_m$, with the earlier assumption $ \omega_m \approx \omega_s$ and
\begin{equation}
   \omega_s \equiv \sqrt{\frac{1}{L_mC_m}},
\end{equation}
the new expression for $C_T$ takes the form \cite{Wu2020}
\begin{equation}\label{eq:CT closed}
    C_T = \frac{C_m}{2} \left[ \sqrt{1+\frac{4}{R_{EM}^{opt}\omega_s^2C_m^2(Z_{tx}+R_L)}}-1 \right] - C_0
\end{equation}
where the matching inductance can now be calculated via
\begin{equation}
     L =\frac{L_mC_m(C_0+C_T)}{(C_0+C_T)(C_m+C_0+C_T)} = \frac{1}{\omega_s^2(C_m+C_0+C_T)}.
\end{equation}
How do these expressions change when we wish to minimize the added noise during transduction? When this is our goal over and above efficiency maximization, we turn away from impedance matching as commonly understood. Here, we no longer wish to achieve zero signal reflection on resonance ($\omega_m = \omega_{LC}$). Instead, we want to maximize the electromechanical cooperativity $\mathcal{C}_{EM}$ to minimize the number of thermal noise quanta (Equation~\ref{nth}). Since  $\mathcal{C}_{EM}$ is inversely proportional to the matching capacitance, we can simply set $C_T = 0$, and calculate the matching inductance $L$ at the resonant condition as is done to maximize conversion efficiency.

Detailed derivations behind this brief summary section can be found in reference \cite{Wu2020}. In Sec.~\ref{matching_results} below, we use this formalism to design impedance matching networks for FBAR transducers. Before detailing these results, we must ask, can the formalism be modified to include an additional cooperativity? If so, how? Such a modification is necessary for transducers that utilize two coupled optical cavities. We explore this question next.

\begin{table}[b]
\caption{\label{tab:blesin}%
Input Parameters for Extended BVD Model.}
\begin{ruledtabular}
\begin{tabular}{lccc}
\textrm{Name}&
\textrm{Symbol}&
\textrm{Value}\\
\colrule
Single photon \\optomechanical coupling rate & $g_{OM,0}/2\pi$ & $400\;\textrm{Hz}$ \\
Optical cavity frequency & $\omega_o/2\pi$ & $193\;\textrm{THz}$ \\
Optical intrinsic linewidth & $\kappa_i/2\pi$ & $25\; \textrm{MHz}$ \\
Optical external coupling rate & $\kappa_{ext}/2\pi$ & $125\; \textrm{MHz}$ \\
Mechanical mode frequency & $\omega_{m}/2\pi$ & $3.285\; \textrm{GHz}$ \\
Mechanical intrinsic linewidth & $\gamma_{i}/2\pi$ & $2.6\; \textrm{MHz}$ \\
Electromechanical coupling factor & $k_{eff}^2$ & $4.3\times10^{-3}$ \\
Optical Cavity Photon Number & $n_{cav}$ & $1\times 10^6 -2.5\times10^9$ \\
Static capacitance & $C_0$ & $200\; \textrm{fF}$ \\
Static resistance & $R_0$ & $10\; k \Omega$ \\
Device Temperature & $T$ & $10-100\; \textrm{mK}$ \\
Ring-ring coupling rate & $J/ 2\pi$ & $1.7 \;\textrm{GHz}$\\
\end{tabular}
\end{ruledtabular}
\end{table}

\begin{figure}
\includegraphics[width =0.45\textwidth]{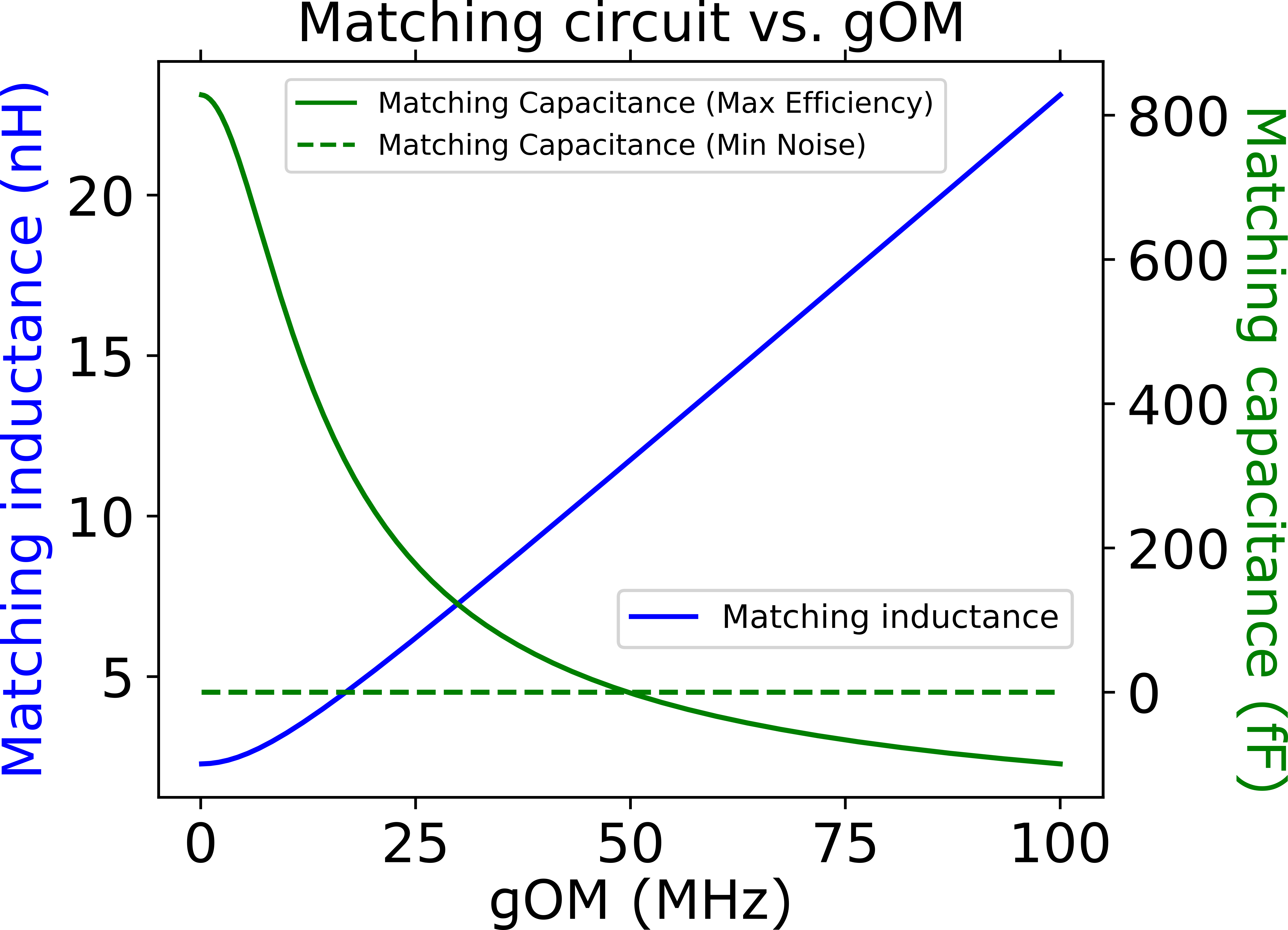}
\caption{Matching inductance (blue) and capacitance (green) for a single-ring FBAR transducer as a function of $g_{OM}$. The matching inductance is the same for efficiency maximization and noise minimization, and increases linearly for $g_{OM}> 3\;\mathrm{MHz}$. For a noise-minimizing circuit, the matching capacitance is constant at zero. For an efficiency-maximizing circuit, the matching capacitance decreases with increasing $g_{OM}$. The formalism outputs a negative (un-physical) matching capacitance for $g_{OM}>50\;\mathrm{MHz}$.}
\label{fig:Fig3m}
\end{figure}

\subsection{Modifying the equivalent circuit to incorporate a photonic molecule}
Here we modify the extended BVD model to include the optomechanical loading from a photonic molecule, i.e. a system of two coupled micro-ring resonators, for the first time. A photonic molecule is a two-level photonic system that can be dynamically controlled by gigahertz-frequency microwave signals, and it typically consists of two coupled MRRs, where only one MMR is coupled to a bus waveguide \cite{Zhang2019}. This extension is a crucial step for modeling a wide range of optomechanical transducers. While single-ring FBAR transducers have demonstrated promising performance for transduction, many leading transducers incorporate double-ring optical resonators. Double-ring resonators enable more straightforward bidirectional operation\cite{blesin2023} and, in some cases, frequency tunability of the hybridized optical supermodes \cite {Fu2021, Warner2025}. Many rely on the tunable mode splitting within a photonic molecule to achieve triply resonant transducer operation \cite{blesin2023}.  Beyond this, the supermode tunability can allow for frequency matching of the emission from two separate transducers, at least on small frequency scales, to correct for modest fabrication variations across separate devices that may impact the degree of in-distinguishability required for interference. Since we aim to incorporate these transducers into interference-based quantum networking protocols, there is an urgent need to generalize the impedance matching formalism to the double-ring class of devices. See Appendices \ref{app:distinguish}, \ref{app:spatiotemporal} for in-depth discussions of these issues.

When two MRRs are coupled at a rate $J$, their modes hybridize into supermodes. So-called symmetric and anti-symmetric supermodes appear with a frequency splitting equal to $2J$. For microwave-to-optical transduction, the optical  pump is tuned to the lower-frequency symmetric supermode frequency $\omega_s$, which sits $\omega_m = 2J$ lower than $\omega_A$. This triply-resonant condition, when satisfied, should boost the conversion efficiency at a given optical power due to the cavity enhancement of the pump tone \cite{blesin2023}.

The coupling between the two MRRs introduces additional optomechanical loading on the mechanical mode, so we must re-define the equivalent optomechanical circuit for this device. We do this by adding two additional resistors $R_{OO,\pm}$ to represent the optical signal and noise outputs, and add these in parallel with a modified optomechanical coupling resistor $R_{OM}$.  $R_{OO,\pm}$ are added in parallel, as opposed to in series, following the generic formalism put forth in Ref. \cite{WangJiang2022}. The addition of a second MRR, and the corresponding optical-optical cooperativity $\mathcal{C}_{OO}$, transforms the device from a 1-stage transducer to a 2-stage transducer. As such, the effective electrical circuit renders the correct transmission coefficient for the device within the Heisenberg-Langevin equations of motion \cite{WangJiang2022}. 

The coupling between the acoustic mode and the first optical cavity is now represented by the effective impedance $R_{OM}$ as 
\begin{equation}
    R_{OM} = R_m  \mathcal{C}_{OM}.
\end{equation}
This term now represents the optomechanical interaction without including information about the output sideband amplitudes. To capture that information, we add in two resistive terms. The resistors $R^{\pm}_{OO}$ are defined as
\begin{multline}
\cr
    R_{OO, +} = R_m  \mathcal{C}_{OO} \mathscr{L}^2_+ \\
    R_{OO, -} = R_m  \mathcal{C}_{OO} \mathscr{L}^2_- \cr
\end{multline}
For the newly defined optical-optical cooperativity $ \mathcal{C}_{OO}$
\begin{equation}
     \mathcal{C}_{OO} = \frac{4J^2}{\kappa_1\kappa_2}
\end{equation}
where we assume the intrinsic linewidths of both cavities are equal, $\kappa_1 = \kappa_2 = \kappa_o$, and $J$ is the ring-ring coupling rate. These terms include the information about Anti-Stokes and Stokes output sideband amplitudes. We note that, while these changes to the equivalent circuit modify the expressions for the optomechanical circuit elements, they do not change the expressions for the piezoelectric sub-circuit or impedance matching network elements.

For analysis of the circuit, we add the optomechanical resistors together in parallel, yielding the expression
\begin{multline}
     R_{OM}^{eq} = \left( \frac{1}{R_{OM}}+\frac{1}{R_{OO,+}-R_{OO,-}}\right) ^{-1}\\
      = R_{OM,+}^{eq}-R_{OM,-}^{eq}\\
      = R_m \mathcal{C}_{OM}^{eq}(\mathscr{L}_+^2-\mathscr{L}_-^2)
\end{multline}
where the modified optomechanical cooperativity is defined as
\begin{equation}
      \mathcal{C}_{OM}^{eq} = \frac{ \mathcal{C}_{OM} \mathcal{C}_{OO}}{ \mathcal{C}_{OM}+ \mathcal{C}_{OO}(\mathscr{L}_+^2-\mathscr{L}_-^2)}.
\end{equation}
And the optomechanical loading on the acoustic mode is now
\begin{equation}
     R_{EM}^{opt} = R_m + R_{OM}^{eq}.
\end{equation}
What expressions should we use to extract the device's figures of merit now? The expression for thermal noise remains unchanged (see Equation~\ref{nth}), but the expressions for Raman noise, conversion efficiency and bandwidth do change. The optical Raman noise $n_o$ is now
\begin{equation}
    n_o = \frac{1}{\eta_e}\frac{\mathcal{C}_{OM}^{eq}\mathscr{L}_-^2}{\mathcal{C}_{EM}},
\end{equation}
the bandwidth is defined as
\begin{multline}
  \Delta \omega = \gamma_m (1+  \mathcal{C}_{EM} +  \mathcal{C}_{OM}^{eq}(\mathscr{L}^2_+-\mathscr{L}^2_-)) \\ = (R_m+R_{EM}+R_{OM}^{eq})/L_m
\end{multline}
so we see that the additional optomechanical loading further broadens the conversion bandwidth. What of the efficiency? Following the standard expression, which applies when $\mathcal{C}_{OM}, \mathcal{C}_{EM}, \mathcal{C}_{OO}<1$, we write
\begin{multline}
    \label{eta_1ring_matched}
   \eta = \eta_e \eta_o \frac{4R_{EM}R_{OM,+}^{eq}}{(R_m+R_{EM}+R_{OM}^{eq})^2}\\
   = \eta_e \eta_o \frac{4 \mathcal{C}_{EM} \mathcal{C}_{OM}^{eq}\mathscr{L}_+^2}{(1+ \mathcal{C}_{EM}+ \mathcal{C}_{OM}^{eq}(\mathscr{L}_+^2-\mathscr{L}_-^2))^2}.
\end{multline}
However, for the two-ring device, we are far outside of the low-cooperativity regime. The strong ring-ring coupling needed to achieve supermode splitting at relevant frequencies of $~4\;\mathrm{GHz}$ results in large $\mathcal{C}_{OO}$, and consequently, the matching circuit yields a large $ \mathcal{C}_{EM}$. For $ \mathcal{C}_{EM}\gg \mathcal{C}_{OM}$, $\mathcal{C}_{OO}\gg\mathcal{C}_{OM}$, we must also consider the alternative expression \cite{Wang2022}:
\begin{multline}
    \label{eta_1ring_matched_alt}
   \eta_{alt} = \eta_e \eta_o \frac{\mathcal{C}_{OM}}{(\sqrt{\mathcal{C}_{OM}+1}+1)^2}\\
    = \eta_e \eta_o \frac{R_{OM}}{(\sqrt{R_{OM}+R_m}+\sqrt{R_m})^2}
\end{multline}
as with the 1-ring transducer, the alternative efficiency expression has no dependence on the matching circuit. We include both the standard and alternative efficiency values in Table \ref{tab:results}.

\begin{figure*}
\includegraphics[width =\textwidth] {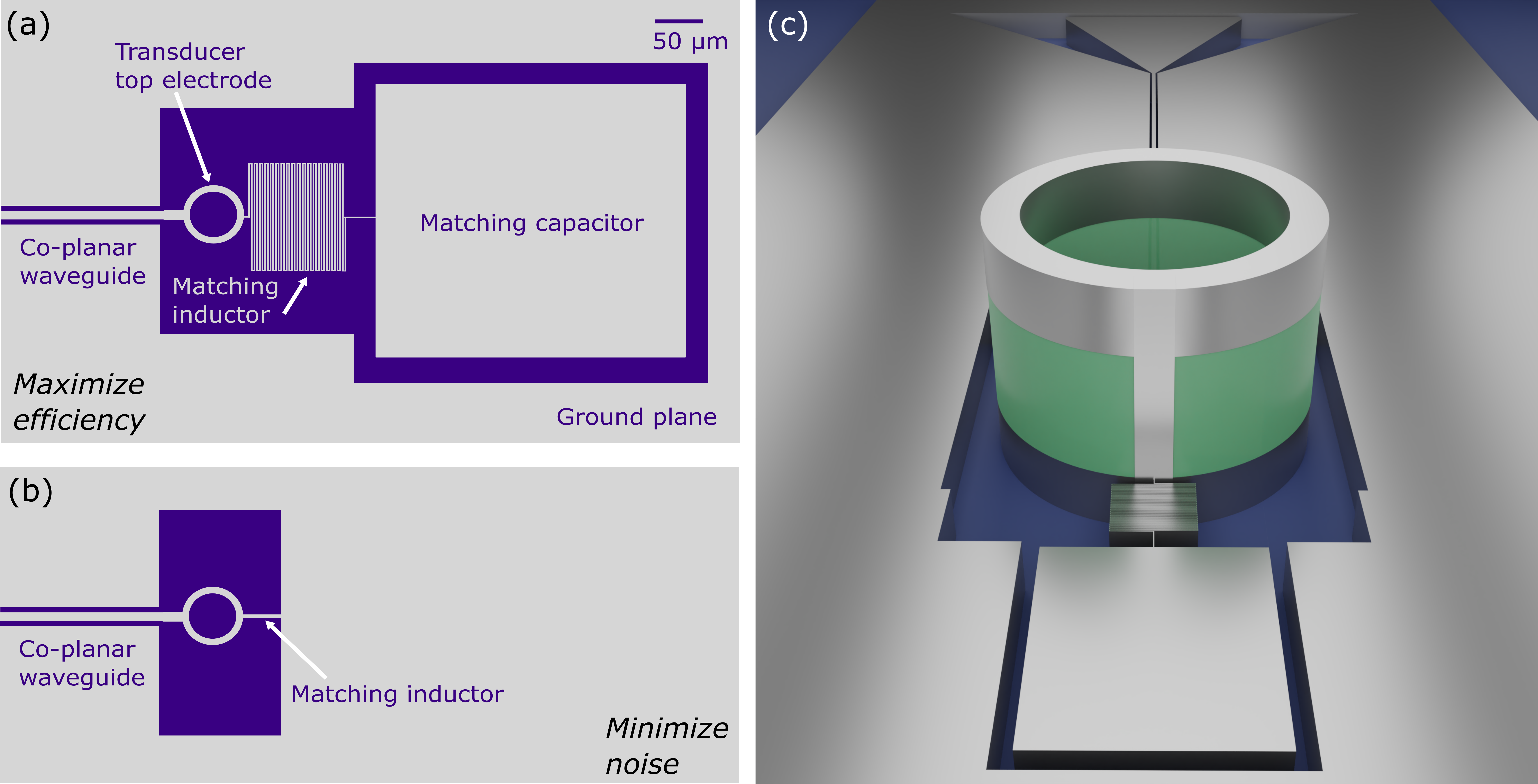}
\caption{Example device designs for a single-ring FBAR transducer. The matching networks shown here are qualitatively designed for an FBAR transducer with the properties listed in Table \ref{tab:blesin} and $g_{OM}=10\;\mathrm{MHz}$. The matching network depicted in (a,b) is designed to maximize the conversion efficiency of the transducer, and consists of a meandering inductor and a square-shaped capacitor. The 3D rendering in (b) shows an electrical wire bonding pad leading to a $50\;\mathrm{Ohm}$ superconducting waveguide, which is galvanically connected to the transducer. The green layer in (b) is the AlN piezo layer. Not shown here is the piezo cavity's bottom electrode and the cladded MRR, sitting beneath the piezo cavity. The device depicted in (c) has a matching network designed to minimize added noise, which consists of a high kinetic inductance nanowire inductor. The design is similar for double-ring devices: a second optical ring resonator would be coupled to the ring located beneath the top electrode shown. Note that (a,b) show a MRR with a radius of $20\;\mathrm{\mu m}$ to match reference \cite{Blesin2021}, whereas (c) has a MRR radius of $125\;\mathrm{\mu m}$ to match reference \cite{Tian2020}.}
\label{fig:Fig4}
\end{figure*}
\subsection{Incorporating parasitic circuit elements}
The formalism developed so far, like previous work, ignores stray capacitances associated with the matching inductor $L$ in the equivalent circuit model. This is despite the large matching inductances and small matching capacitances often required for OMC-based transducers \cite{Wu2020}, which may require large spiral inductors that have large stray, or parasitic, capacitances. The stray inductance of the matching capacitor should also be considered. As explored in-depth in Appendix \ref{app:parasitic}, we derive updated expressions for the matching capacitance and matching inductance that account for parasitic terms. The modified matching terms ($C_T’$, $L’$) can be written as:

\begin{equation}
C_T' = C_T + C_p
\end{equation}
and
\begin{equation}
L' = L + L_{CT}.
\end{equation}

\begin{figure}
\includegraphics{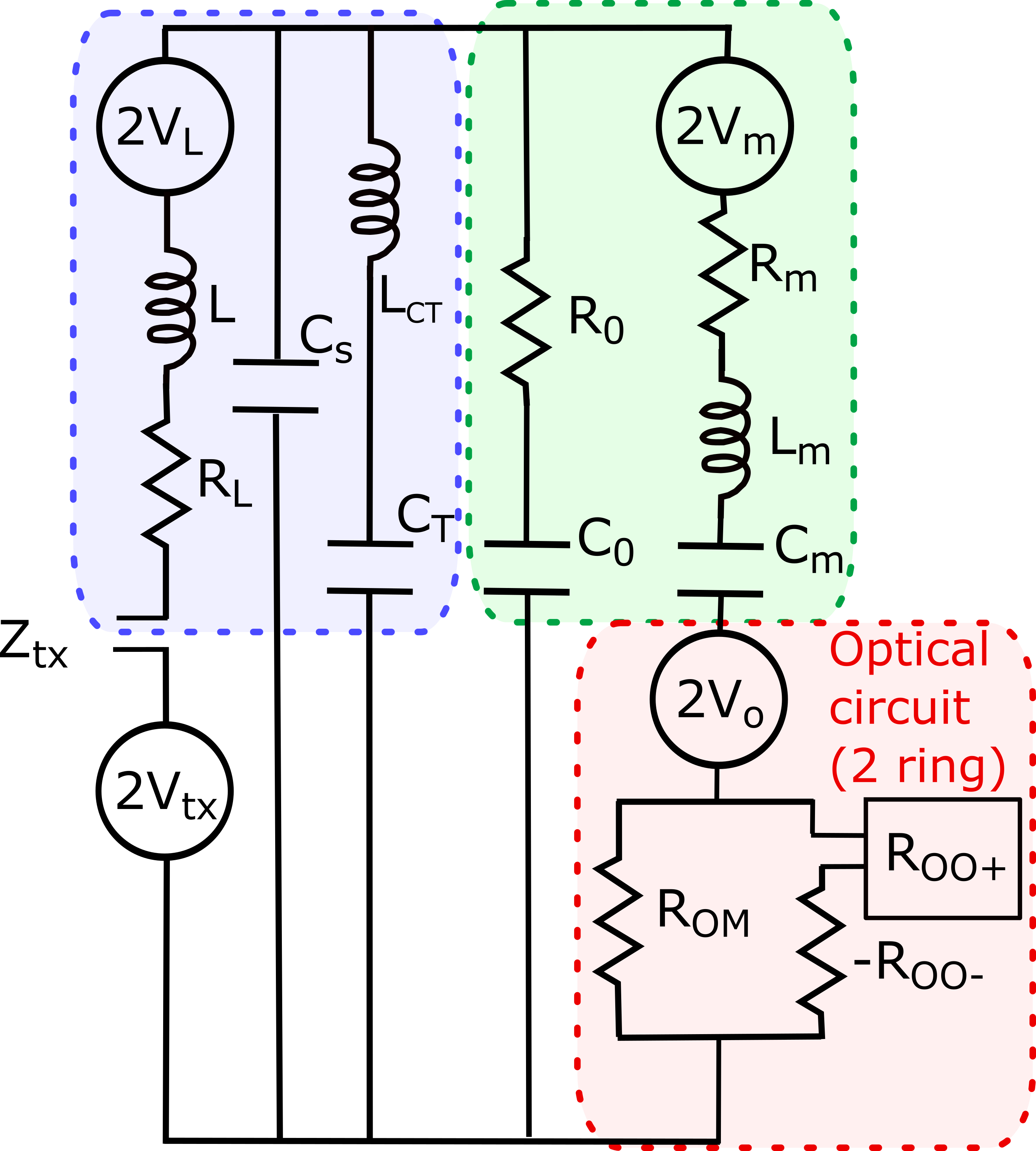}
\caption{A modified equivalent circuit for a transducer with parasitic terms incorporated into the matching network. $L_{CT}$: parasitic inductance of the matching capacitor. $C_s$: parasitic capacitance of the matching inductor.}
\label{fig:Fig11}
\end{figure}

An updated equivalent circuit diagram incorporating elements for parasitic terms is shown in Fig. ~\ref{fig:Fig11}.

\subsection{\label{matching_results}Matching circuits for single-ring and double-ring FBAR transducers}
Here we apply the above formalism to the design of matching circuits for FBAR transducers. For all calculations, we use physical parameters from Blesin \emph{et al.}\cite{Blesin2021} as inputs to the model: these parameters are listed in Table~\ref{tab:blesin}.  For all calculations, we also set the device temperature to $50\;\mathrm{mK}$ and the cavity-enhanced optomechanical coupling rate to $10\;\mathrm{MHz}$, which corresponds to $6.25 \times 10^8$ photons in the optical cavity. The frequency of the optical pump is set to be red-detuned from the optical cavity mode by the acoustic mode frequency: $\Delta_o = -\omega_m$. Results obtained for single- and double-ring devices are summarized in Table \ref{tab:results} and discussed briefly here.

\subsubsection{Matching Circuit for Single-Ring FBAR Transducer}
The matching capacitance and inductance values required for a single-ring FBAR transducer are plotted versus $g_{OM}$ in Figure \ref{fig:Fig3m}. The plot shows how the required matching capacitance increases versus pump power for an efficiency-maximizing circuit. At $g_{OM}>50\;\mathrm{MHz}$, the matching capacitance becomes negative. This un-physical result tells us that it is not possible to design an efficiency-maximizing impedance matching circuit at this pump power. The matching inductance is equal for either efficiency maximization or noise minimization, and increases with pump power. Overall, this plot shows that it is easier to fabricate matching circuits for lower pump power operation of the transducer. 

An impedance matching circuit designed to maximize conversion efficiency at a moderate pump power with $g_{OM}=10\;\mathrm{MHz}$ requires a matching capacitance of $522.346\;\mathrm{fF}$ and a matching inductance of $L= 3.246\;\mathrm{nH}$. This circuit results in a conversion efficiency of $42.197\%$. Since all cooperativities have values close to $1$, the standard efficiency expression matches the alternative efficiency expression. The conversion bandwidth is $10.533 \; \mathrm{MHz}$, the Raman noise quanta are $6.598 \times 10^{-5}$ per transduced photon, and the thermal noise quanta are $2.203-2 \times 10^{-2} $ per transduced photon.

When we modify the matching circuit to minimize added noise, $C_T \rightarrow 0 \;\mathrm{fF}$, the required matching inductance remains almost the same. With this matching circuit, we achieve a conversion efficiency of $28.688\%$ (standard expression). The decrease in efficiency is to be expected, since the matching circuit no longer minimizes signal reflections when $\omega_{LC}=\omega_m$. We add  $1.829\times 10^{-5}$ Raman noise quanta per transduced photon, and $6.110\times 10^{-3}$ thermal noise quanta per transduced photon. Choosing this circuit, we can lower the thermal noise quanta by $27.7\%$. In the process, we sacrifice conversion efficiency. Using the standard efficiency expression,the efficiency drops -  however, $\mathcal{C}_{EM}=7.304>1$, outside of the low-cooperativity regime where the standard expression applies - so we must be wary of this value. The alternative efficiency value ($42.197\%$) doesn't change, since it does not depend on matching circuit parameters. We can then understand the peak conversion efficiency to sit in a band between these upper and lower bounds, understanding the efficiency is likely to be lower than that for the efficiency-maximizing matching network.
\begin{figure*}
\includegraphics[width =\textwidth] {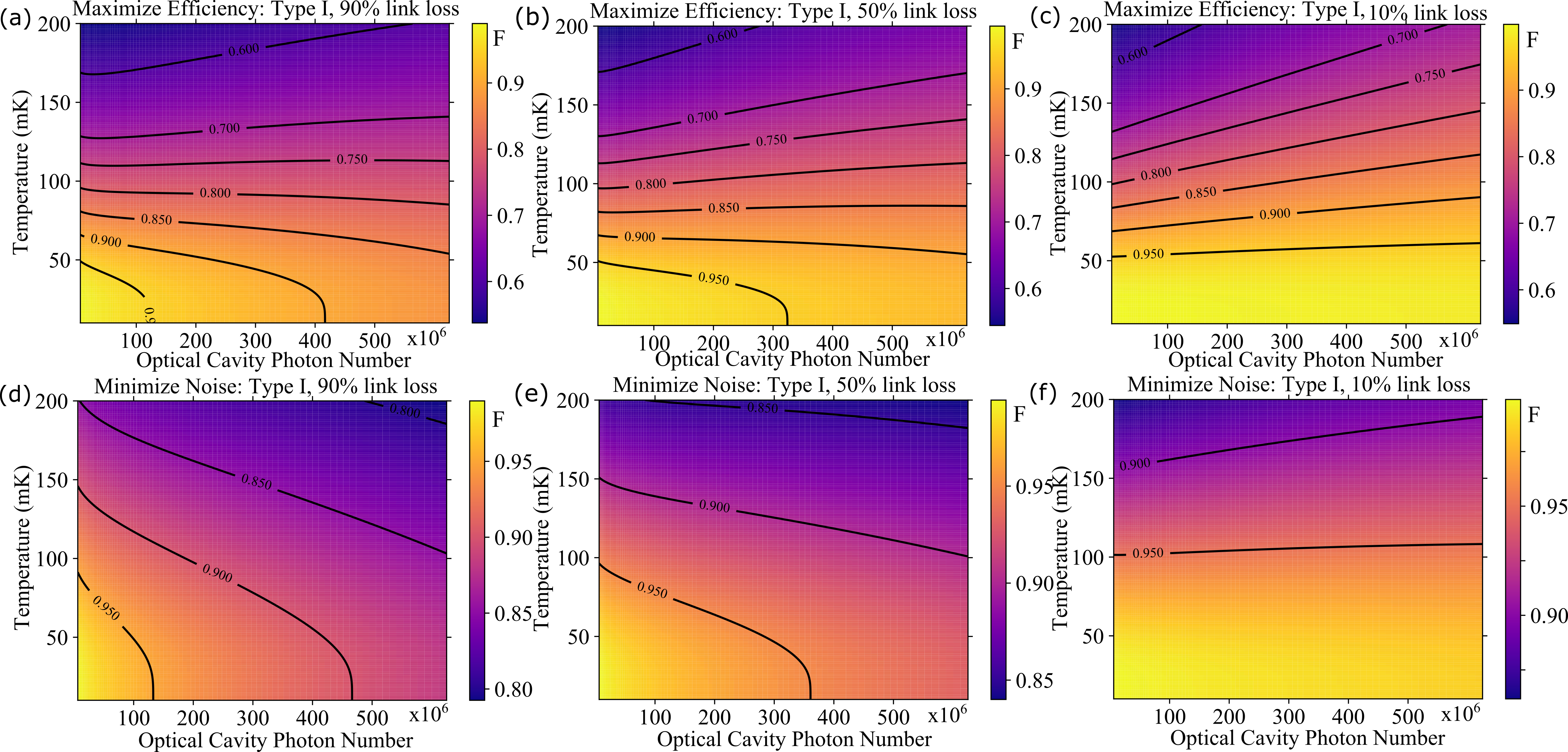}
\caption{Entanglement readout fidelity F for Type-I heralding protocols plotted versus optical cavity photon number and device temperature. The plots are shown with link loss values of $90\%$ (a,b), $50\%$ (c,d) and $10\%$ (e,f). These results show that the achievable readout fidelity has a strong dependence on $\eta_l$, which is mitigated to a small degree using noise-minimizing matching circuits. Overall, the achievable fidelity is boosted by at least $5\%$, and up to $10\%$ or more, at higher temperatures when the noise-minimizing matching circuit is used.}
\label{fig:Fig5}
\end{figure*}

The optimal matching circuit values depend sensitively on the operating regime of the transducer. For example, for $g_{OM} = 10\;\mathrm{MHz}$, the matching capacitance to optimize conversion efficiency is $522\;\mathrm{fF}$. If we increase the number of optical pump photons from $6.25 \times 10^8$ to  $2.5 \times 10^9$, a four-fold increase, the optimal matching capacitance falls to $253\;\mathrm{fF}$. These results reinforce the  requirement to design and fabricate transducers to operate within a specific set of experimental conditions, informed by the resulting estimated figures of merit entanglement protocols, as we obtain below. See Fig.~\ref{fig:Fig4} for an example impedance-matched transducer design. 

When designing impedance matching networks, especially those that require $C_T=0$ to minimize added noise, we must take care to consider the effects of  parasitic capacitance terms. Spiral inductors, like the one depicted in Ref. \cite{Wu2020}, have significant self-capacitances and capacitances to ground. If ignored, stray capacitances will pull the matching circuit away from its optimal regime, leading to degraded device performance. Likewise, matching capacitors will have parasitic inductances. In Appendix \ref{app:parasitic} we investigate the effects of these unwanted terms on the matching network and transducer performance.

\subsubsection{Matching Circuit for Double-Ring FBAR Transducer}
For a two-ring transducer, the matching circuit must be designed to match the new optomechanical sub-circuit. Of note, adding the new resistors $R_{OO,\pm}$ in parallel with $R_{OM}$ only changes the total optomechanical loading very minimally, from $90.314\;\mathrm{\Omega}$ to $90.229\;\mathrm{\Omega}$. Because of the similar optomechanical loading, the matching circuit values and corresponding figures of merit are similar for both instantiations of the device, both for efficiency-maximizing and noise. The only notable difference is that the alternative efficiency expressions for the two-ring device return a lower efficiency value. It appears that for the single ring devices, the appropriate standard expression overestimates the conversion efficiency, whereas for the two-ring device, the appropriate standard expression underestimates the efficiency.

While the matching circuits for one-ring and two-ring transducers are similar for the operating regime studied here, we are now equipped with a more generalized formalism that enables us to design matching networks across a wider variety of optomechanical transducers. One such design is discussed below.
\begin{figure}
\centering
\includegraphics[width=.45\textwidth]{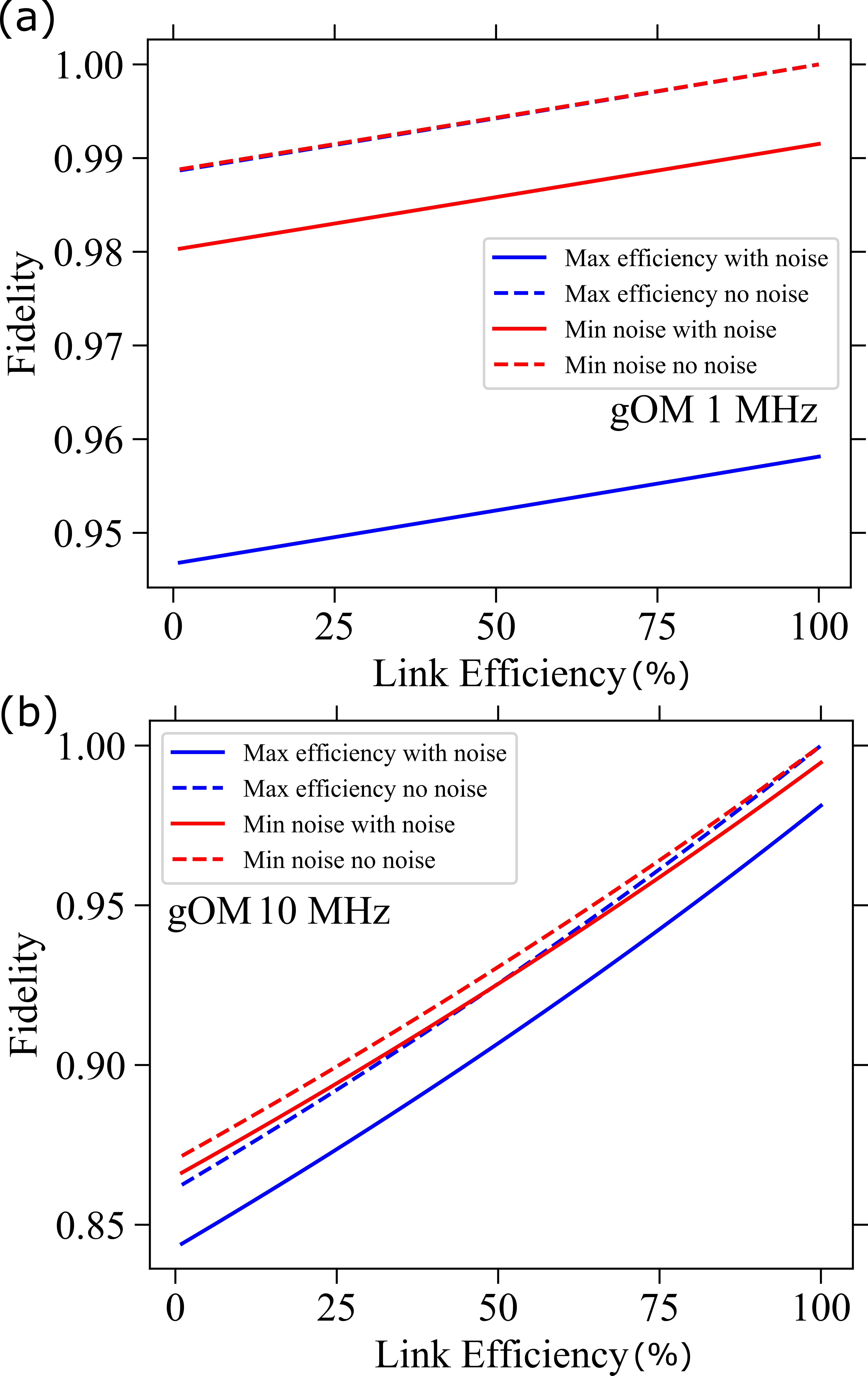}
\caption{Type-I Entanglement readout fidelity versus link loss for $T =50 \;\mathrm{mK}$  and (a) $g_{OM}=1 \; \mathrm{MHz}$ and (b) $g_{OM}=10 \; \mathrm{MHz}$. Blue curves correspond to efficiency-maximizing matching circuits and red curves correspond to noise-minimizing matching circuits. Dashed curves correspond to fidelity ignoring the effects of thermal noise, and solid curves take thermal noise into account. For larger $g_{OM}$, the difference in fidelity between the two matching circuits is more prominent, since the probability of false positive detection events increases more dramatically for high link loss.}
\label{fig:Fig6}
\end{figure}
\begin{figure}
\centering
\includegraphics[width=0.45\textwidth]{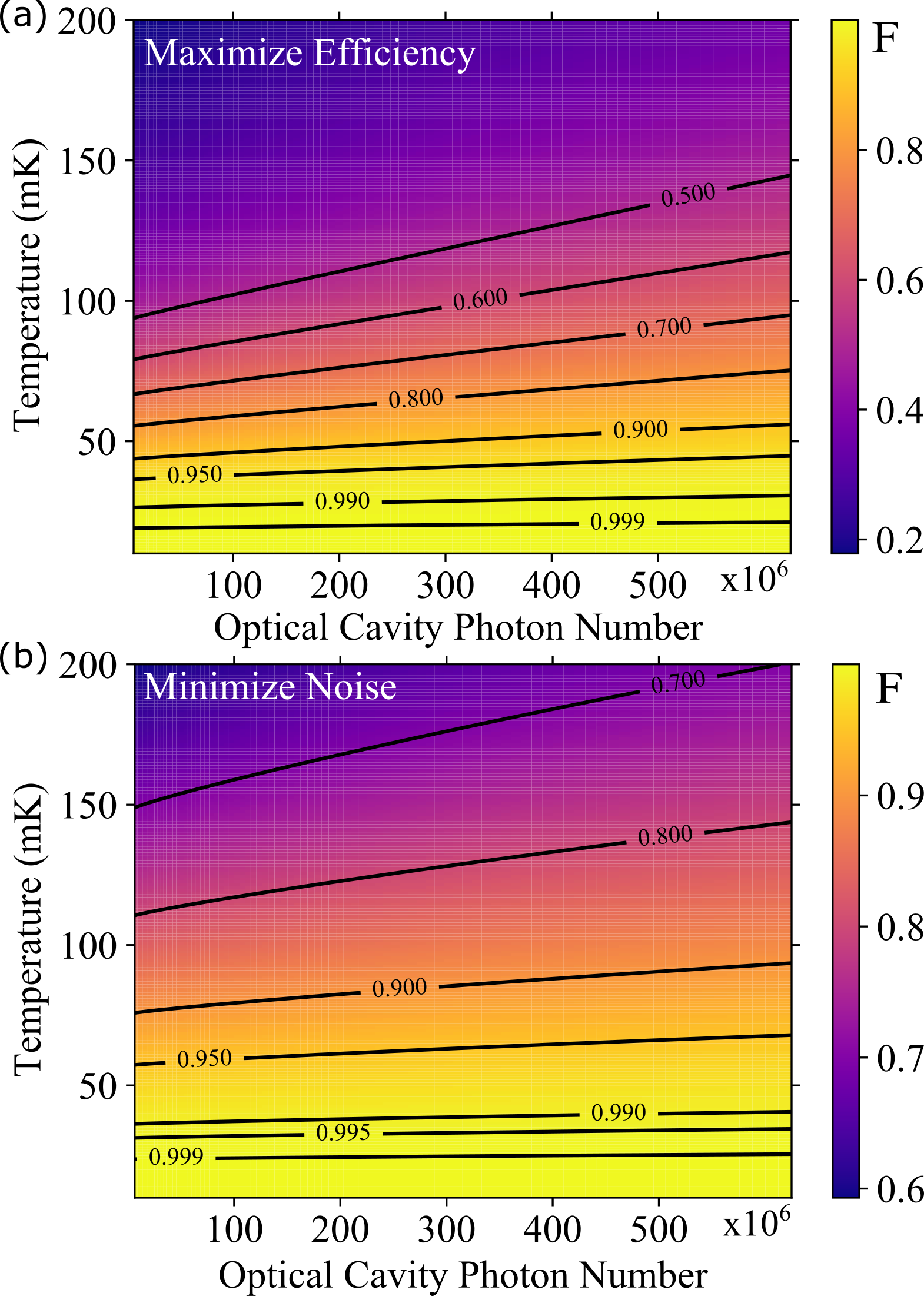}
\caption{Type-II Entanglement fidelity F versus temperature and optical cavity photon number for a matching circuit (a) maximizing conversion efficiency and (b) minimizing added noise. The Type-II protocol is not susceptible to lower fidelities at higher pump powers, as opposed to the Type-I protocol. The addition of a noise-minimizing impedance matching circuit in (b) significantly improves fidelity as the temperature increases.}
\label{fig:Fig7}
\end{figure}

\subsubsection{Example Matching Circuit Design}
Using the above formalism, we here propose device designs for FBAR transducers with both efficiency-maximizing and noise-minimizing matching circuits. Rough order-of-magnitude qualitative designs and renderings for these devices are shown in Fig.~\ref{fig:Fig4}, where the FBAR transducer is understood to have the material stack shown in the cross-section in Fig.~\ref{fig:Fig1}(a). The device shown here has a MRR with a radius of $20\;\mathrm{\mu m}$, ($125\;\mathrm{\mu m}$ in panel (c)), and an annulus-shaped top electrode. The electrode is fabricated using a pull-back process \cite{Siddharth2024, blesin2023} engineered to minimize electrode capacitance.  This is a qualitative diagram, and the design is understood to be applicable to a variety of optical cavities and electrode configurations.

The example designs are shown with a $200\;\mathrm{nm}$ thick Niobium superconducting layer, which defines the superconducting circuit and the ground plane. A wire-bonding pad (top of panel (b)) connects to a $50\;\mathrm{Ohm}$ superconducting waveguide, which connects galvanically to the top electrode of the transducer. This electrode is also galvanically connected to the impedance matching circuit to achieve the in-parallel configuration dictated by the extended BVD circuit model. The ground plane for the impedance matching circuit is the same as the ground of the transducer's piezo cavity. The efficiency-maximizing version of the impedance matching network shown in panel (a,b) incorporates a Niobium meandering inductor.  The total footprint is an area of approximately $100 \;\mathrm{\mu m} \times 100 \;\mathrm{\mu m}$. The meander line width, center-to-separation, and total length should be designed to target a total inductance in the $2-4\;\mathrm{nH}$ range with an anticipated parasitic capacitance in the $5-10\;\mathrm{fF}$ range.  The specific lumped element values will depend on the quality of the deposited materials and fabrication conditions, so this example is meant to be qualitative and not exact. The matching capacitor is designed in a rectangular shape with dimensions of approximately $300 \;\mathrm{\mu m} \times 325 \;\mathrm{\mu m}$, and has an additional capacitance of $400-500\;\mathrm{fF}$.

A device with a noise-minimizing version of the matching network is shown in panel (c). Since this matching network requires a capacitance as close to zero as possible, the square capacitor is discarded and the meandering inductor is swapped out for a high kinetic inductance nanowire inductor \cite{Niepce2019}. The nanowire inductor shown consists of a $\sim20\;\mathrm{nm}$-thick 
NbN layer with a width of $1\;\mathrm{\mu m}$ and length of $\sim150 \;\mathrm{\mu m}$. Depending on the specific materials and fabrication recipes involved, the nanowire length can possibly be reduced to $<10\;\mathrm{\mu m}$ to achieve an inductance in the $2-4\;\mathrm{nH}$ range with an estimated parasitic capacitance is in the $\mathrm{aF}$ range. Additional discussion exploring the effects of parasitic lumped element terms can be found in Appendix \ref{app:parasitic}. The impacts of fabrication variances of the matching circuit elements on transducer and entanglement heralding performance are discussed in Appendix \ref{app:sensitivity}. A sensitivity analysis reveals that the FBAR transducers are robust to realistic fabrication tolerances, as are the quantum networking protocols that rely on these devices.

\section{\label{Entanglement Heralding}BVD-Informed Entanglement Heralding Protocols}

Now that we have established how to design impedance matching networks for the FBAR transducer, both to maximize conversion efficiency and minimize added noise, we investigate how the impedance-matched devices perform within Type-I and Type-II entanglement heralding protocols. The generalized extended BVD model developed in Section II provides guidance for our current task by connecting device-level optimization to protocol-level performance. The above formalism enables systematic selection of optimal transducer operating points that balance conversion efficiency and added noise for maximum entanglement fidelity. As we will show, this framework allows quantum network engineers to make informed trade-offs between entanglement generation rate and fidelity based on application requirements.

Toward this end, conditional readout fidelities and entanglement generation rates are plotted and analyzed in various experimental operating conditions. These results allow us to make informed decisions in the device design phase to optimize the chosen quantum networking protocol, given a set of realistic experimental parameters such as temperature, transducer pump power, and link loss between the transducer and the Bell state analyzer.

Here we define the conditional readout fidelity $F$ as the overlap between the generated quantum state and the target Bell state, given that the proper heralding click condition is satisfied:

\begin{equation} \label{eq:fid}
F= \bra{\Psi_{target}}\rho_{generated}\ket{\Psi_{target}}
\end{equation}

where $\rho_{generated}$ is the density matrix of the state conditioned on successful heralding, and $\ket{\Psi_{target}=\ket{\Psi^{\pm}}}$ is the desired Bell state \cite{Duan2001, Krastanov2021}.

Let us begin with Type-I entanglement heralding. Because this is a single-photon protocol, it will succeed only when one photon is emitted, in total, from both network nodes (see Fig.~\ref{fig:Fig2}). For each trial of the experiment, there is a small chance that both superconducting network nodes will produce an entangled telecom photon, and that one of these photons will be lost in the network before reaching the photodetector. Instances such as these, where two photons are emitted but only one is detected, and the two events are completely uncorrelated, produce false positive events and contribute to the infidelity (i.e., the decrease in fidelity). In classifying false positive detection events this way, we are assuming that two-photon detection events can be sorted from one-photon detection events and discarded through post-selection.

While this source of infidelity is limiting, in practice, the primary downside of Type-I entanglement heralding is the requirement of absolute relative phase stability between the photon sources and the beamsplitter \cite{Luo2009}. Slight changes in optical fiber path length due to temperature fluctuations, for example, can change the relative phase of the entangled state. This approach could still be suitable for entanglement generation between cryostats in the same building, e.g. local area networks, but will quickly become untenable for longer networks with larger inherent phase instabilities. In the discussion below, the fidelity, or readout fidelity, of the protocol refers to the conditional fidelity of Bell state generation \cite{Zeuthen2020}, that is, the overlap of the created quantum state with the desired quantum state, given that the relevant photodetector click condition is fulfilled. 

Wang \emph{et al.} developed a framework to extract the fidelity and entanglement generation rate of a heralding protocol that depends on the properties of an optomechanical transducer\cite{Wang2022}, but did not incorporate contributions from added noise. Zeuthen \emph{et al.} developed expressions for entanglement fidelity in the presence of noise \cite{Zeuthen2020} and found that the fidelity depends much more sensitively on the number of added noise quanta than on conversion efficiency. Here, we combine the methods of \cite{Wang2022, Zeuthen2020} while establishing a connection to the extended BVD model for an impedance-matched transducer.

Assuming a Poissonian photon detection process, following the Stochastic master equation \cite{Krastanov2021}, a photon is emitted by the transducer with approximate probability 
\begin{equation} \label{eq:poisson_p1}
    P_1 \approx 1- P_0 = 1- e^{-r_o \Delta t}
\end{equation}
where $r_o$ is the photon generation rate and $\Delta t$ is the characteristic time duration of the transduction process. Note that Eq.~\ref{eq:poisson_p1} represents the probability that \emph{at least} one photon is emitted, and not precisely one photon. The probability of precisely one photon emission event occurring within $\Delta t$ is 
\begin{equation} \label{eq:poisson_one_photon}
    P_{single} =  r_o \Delta t e^{-r_o \Delta t}.
\end{equation}
$P_1 \approx P_{single}$ when $P_1 \ll 1$, i.e. when the product $r_0 \Delta t < 1$. As will be shown below (see Table \ref{tab:results2}), we obtain the values at or near $r_0 \Delta t = 0.28$ at a relatively high optomechanical coupling rate of $10\;\mathrm{MHz}$. For this value, $P_1 \approx P_{single}$. We must remain aware of this assumption, however, understanding that it becomes less accurate as the pump power to the transducer is increased.

Next we define the expression for $\Delta t$. For an FBAR transducer with a single optical cavity, we have the characteristic transduction time duration
\begin{equation}
    \Delta t = 2 \pi \left( \nicefrac{1}{g_{EM}} + \nicefrac{1}{g_{OM}} + \nicefrac{1}{\kappa_{ext}} \right)
\end{equation}
which can be understood as the sum of the lifetimes of each of the constituent cavity modes of the device, including the time it takes for a transducer photon to couple out of the MRR and into the optical bus waveguide. For a two-MRR transducer we have
\begin{equation}
    \Delta t_{2MRR} = 2 \pi \left( \nicefrac{1}{g_{EM}} + \nicefrac{1}{g_{OM}} + \nicefrac{1}{J}+ \nicefrac{1}{\kappa_{ext}} \right).
\end{equation}
Both single-MRR and double-MRR devices have an optical photon generation rate (for a single microwave photon input) of
\begin{equation}
    r_o = \frac{1}{2\pi}\frac{4g^2_{OM} \kappa_{ext}}{(\kappa_{ext} + \kappa_i)^2}
\end{equation}
where $\kappa_{ext}$ is the external coupling rate of the optical cavity, and $\kappa_i$ is its intrinsic loss rate. 
The probability of successfully detecting an emitted photon is 
\begin{equation}
    P_{succ} = P_1 \eta_{l} \eta_{det}.
\end{equation}
Where $\eta_{l}$ is the probability that the photon reaches the photodetector without being lost in the network link, and $\eta_{det}$ is the detector efficiency.

The desired, or target, photonic Bell state is 
\begin{equation} \label{eq:target_photonic}
    \ket{\Psi_{phot}}=\sqrt{P_{01}}\ket{0_A1_B} \pm\sqrt{P_{10}}\ket{1_A0_B}
\end{equation}
where $P_{01}$ ($P_{10}$) is the probability that Node A(B) emits zero photons and Node B(A) emits one photon
\begin{equation}
    P_{01}=P_{10} = P_1 P_0 = (1- e^{-r_o \Delta t})e^{-r_o \Delta t}.
\end{equation}

The detection of the target photonic Bell state after the beamsplitter will project the remote superconducting circuits into the distributed Bell state
\begin{equation}
    \ket{\Psi_{SQ}}=\sqrt{P_{01}}\ket{e_Ag_B} \pm \sqrt{P_{10}}\ket{g_Ae_B}
\end{equation}
where $\ket{g} \; , \; \ket{e}$ are the qubits' ground and excited states, respectively. During an entanglement generation attempt, there is a probability that both network nodes will emit entangled photons, but one photon is lost before reaching the photodetector. This scenario contributes false positive events to the infidelity of the protocol, since both superconducting qubits will be in their ground states, even though we think one is in its excited state. This occurs with $P_{in}$, the probability of an infidelity-causing event, of approximately
\begin{equation}
    P_{in} = P_{11}\eta_l(1- \eta_{l}) = P_1^2\eta_l(1 - \eta_{l}).
\end{equation}
As such, the fidelity of the protocol following Equation \ref{eq:fid} is then:
  \begin{multline}\label{fid T1 no noise} \cr
    F = \frac{P_{01}\eta_{l}+P_{10}\eta_{l}}{P_1^2\eta_l(1-\eta_{l})+ P_{01}\eta_{l}+P_{10}\eta_{l}}
    = \frac{2P_{01}}{P_1^2(1-\eta_l)+ 2P_{01}}. \cr
 \end{multline}
In other words, the fidelity is equal to the ratio of the total probability of creating the desired state to the sum of the probabilities of creating the desired state and of creating an undesired state with a single detection event. As the pump power to the transducers increases, $P_1^2$ increases, and the fidelity is degraded. Note that by counting false positive events in this manner, we assume that we are able to distinguish between single count events (desired) and coincident events (undesired) and discard those through post-selection. The associated photon generation rate $n_r$ is 
  \begin{equation}\label{eq: 1 photon gen rate}
    n_r = 2 r_0 e^{-r_0 \Delta t} \left( \frac{\Delta t}{\Delta t + t_r} \right)
 \end{equation}
for a superconducting qubit reset time $t_r$. The factor of $2$ is present because either node can emit a photon. This leads to an entanglement generation rate of  
 \begin{equation}\label{ent gen rate}
    \tau_{ent} = n_r \eta_{l} \eta_{det}.
 \end{equation}
This analysis assumes that the achievable readout fidelity is not reduced by spatio-temporal mode mismatch between the photons emitted by the two nodes. As explored in Appendix \ref{app:spatiotemporal}, within expected fabrication tolerances, we anticipate the spatio-temporal overlap from photons emitted by nominally identical transducers to achieve fidelity ceilings of $>99\%.$

\subsection{\label{Type I}Incorporating Infidelity Contributions Due to Thermal Noise}

For a thermal population of $n_{th}$, the probability of the transducer emitting a photon due to this thermal occupation is approximately
 \begin{equation}
     P_{th} = 1- e^{-r_{th} \Delta t}
 \end{equation}
where $r_{th}$ is the rate of photon generation due to the presence of thermal quanta in the transducer's acoustic mode:
\begin{equation} \label{eq:rth}
     r_{th} = r_o n_{th}.
 \end{equation}

The expression for the infidelity is now
\begin{multline} \label{typeI_infidelity}
     P_{in} = [P_{1}^2(1-P_{th})^2 + \\ P_{01}P_{th}(1-P_{th}) + \\P_{10}P_{th}(1-P_{th})+\\ P_{00}P_{th}^2)]\eta_l(1-\eta_l) + \\2P_{th}(1-P_{th})P_{00}\eta_l.
\end{multline}
The $(P_{01}P_{th}(1-P_{th}) + P_{10}P_{th}(1-P_{th}))\eta_l(1-\eta_l)$ terms correspond to false positive events arising from attempts where one node emits a photon entangled with a superconducting qubit, and the other node emits a photon due to the presence of thermal quanta in the transducer, and one of the photons is lost before reaching the photodetectors. The $P_{00}P_{th}^2\eta_l(1-\eta_l)$ term corresponds to false positive events arising from attempts where both nodes emit photons due to the presence of thermal quanta in the transducer and one of the two emitted photons is lost before hitting the photodetector. There is an additional term of $2P_{th}(1-P_{th})P_{00}\eta_l$ which represents false positive events arising from attempts where neither node emits an entangled photon, but one node emits a photon due to thermal excitation, and that photon reaches a photodetector.

If nodes A and B emit photons with the same probability ($P_{01}=P_{10}$), this simplifies to 
  \begin{multline} \label{typeI_infidelity_s}
    P_{in} = [P_{1}^2(1-P_{th})^2 + 2P_{01}P_{th}(1-P_{th}) + P_{00}P_{th}^2]\eta_l(1-\eta_l) + \\2P_{00}P_{th}(1-P_{th})\eta_{l}
 \end{multline}
The fidelity expression takes the form
\begin{widetext}
\begin{multline}
    F = \frac{2P_{01}(1-P_{th})^2\eta_l}{P_{in} + 2P_{01}(1-P_{th})^2\eta_l}\\
    =\frac{2P_{01}(1-P_{th})^2}{(P_{1}^2(1-P_{th})^2 + 2P_{01}P_{th}(1-P_{th}) + P_{00}P_{th}^2)(1-\eta_l) + 2P_{00}P_{th}(1-P_{th})+ 2P_{01}(1-P_{th})^2}.
\end{multline}
\end{widetext}

For the interested reader, a detailed breakdown of Equations \ref{typeI_infidelity}, \ref{typeI_infidelity_s} is available in Appendix \ref{app:infidelity}. Here is the point where the extended BVD model of our transducer comes into play. For a given set of device and operating parameters, the BVD model outputs matching circuit parameters, an estimated conversion efficiency, and estimates for added noise quanta due to Raman and thermal noise. (Note that in this section we disregard Raman noise quanta, since there are multiple orders of magnitude fewer noise quanta due to Raman noise.) We can use the thermal noise output (Equation~\ref{nth}) from the BVD model as an input to our entanglement generation fidelity expressions, while simultaneously optimizing our transducer for either highest-efficiency or lowest-noise operation. To do this, we input the BVD value of $n_{th}$ into Equation~\ref{eq:rth}.

\subsection{\label{Type II}Two-Photon Protocol with Thermal Noise}


Type-II entanglement heralding involves the interference of two optical photons, one emitted from each node. The interferometric phase is now common to both photons. As such, it can be factored out of the distributed entangled state, bypassing strict requirements on phase stability. Not only are Type-II entanglement protocols more robust to instabilities due to path-length variations, they are also not limited by the intrinsic contribution to the infidelity described for Type-I above, where two photons are emitted but one is lost. This means that the quantum memories can be pumped harder to emit photons with a higher probability. 

For these protocols, the quantum information can be stored in the frequency, state of polarization, or time-bin degree of freedom of the optical photon \cite{Luo2009}. Since superconducting qubits will most naturally emit time-bin encoded photons \cite{kurpiers2019, Ilves2020}, we focus on the time-bin encoding basis for qubit-photon entanglement. Time-bin encoding is commonly found in color-center based quantum networking experiments \cite{Tchebotareva2019, Bersin2024}.

Here, for example, detecting a photon in the Early time bin mode corresponds to a qubit in its ground state, whereas a photon in the Late time bin mode corresponds to a qubit in its excited state. Changing to the time-bin basis, the target photonic Bell state is now
\begin{equation}
    \ket{\Psi_{phot}}=\sqrt{P_{A,L}P_{B,E}}\ket{0_E1_L} \pm\sqrt{P_{A,E}P_{B,L}}\ket{1_E0_L}
\end{equation}

where $P_{A,L}$ is the probability that Node A emits a photon in the Late time bin and $P_{B,E}$ is the probability that Node B emits a photon in the Early time bin. If this photonic state is detected, it will project the superconducting qubits into the target Bell state
\begin{equation}\label{eq:SQ_Bell}
    \ket{\Psi_{SQ}}=\sqrt{P_{eg}}\ket{eg} \pm \sqrt{P_{ge}}\ket{ge}
\end{equation}
where $P_{eg}=P_{A,L}P_{B,E}=P_{1}^2$ is the probability that the qubit at Node A is in the excited state and the qubit at Node B is in the ground state.
\begin{table*}[htpb]
\caption{\label{tab:results}%
Example results for FBAR transducers (for $n_{cav} = 6.25 \times 10^8$ or $g_{OM} = 10\; \mathrm{MHz}$ and $\Delta = \omega_s$ at $50\;\mathrm{mK}$).}
\begin{ruledtabular}
\begin{tabular}{lccccr}
\textrm{Name}&
\textrm{Symbol}&
\textrm{1 Ring, Max Eff.}&
\textrm{1 Ring, Min Noise}&
\textrm{2 Rings, Max Eff.}&
\textrm{2 Rings, Min Noise}\\
\colrule
Motional capacitance & $C_m$& $0.860 \; \textrm{fF}$ & $0.860 \; \textrm{fF}$  & $0.860 \; \textrm{fF}$ & $0.860 \; \textrm{fF}$\\
Motional inductance & $L_m$& $2.729 \; \mathrm{\mu H}$ & $2.729 \; \mathrm{\mu H}$ &$2.729 \; \mathrm{\mu H}$ & $2.729 \; \mathrm{\mu H}$ \\
Motional resistance & $R_m$& $44.588 \; \mathrm{\Omega}$ & $44.588 \; \mathrm{\Omega}$ & $44.588 \; \mathrm{\Omega}$ &$44.588 \; \mathrm{\Omega}$\\
\colrule
Lorentzian upper sideband & $\mathscr{L}_+^2$& $1$ & $1$  & $1$ & $1$\\
Lorentzian lower sideband & $\mathscr{L}_-^2$& $1.303\times10^{-4}$ & $1.303\times 10^{-4}$ & $1.303\times 10^{-4}$ & $1.303\times 10^{-4}$ \\
Upper sideband resistance 1 ring& $R_{OM,+}$& $45.732\;\mathrm{\Omega}$ & $45.732\;\mathrm{\Omega}$  & N/A & N/A\\
Lower sideband resistance 1 ring& $R_{OM,-}$& $5.959\times10^{-3}\;\mathrm{\Omega}$ & $5.959\times 10^{-3}\;\mathrm{\Omega}$ & N/A&N/A \\
Equivalent OM resistance 2 ring& $R_{OM}^{eq}$& N/A & N/A & $45.656\;\mathrm{\Omega}$ & $45.656\;\mathrm{\Omega}$\\
Upper sideband resistance 2 ring & $R_{OO,+}$ & N/A & N/A  & $22.908\;\mathrm{k\Omega}$ & $22.908\;\mathrm{k\Omega}$\\
Lower sideband resistance 2 ring & $R_{OO,-}$& N/A & N/A & $2.985\;\mathrm{\Omega}$ & $2.985\;\mathrm{\Omega}$  \\
Optomechanical cooperativity & $C_{OM}$ & $1.026$ & $1.026$  & $1.026$ & $1.026$ \\
Optical-optical cooperativity & $C_{OO}$ & N/A & N/A  & $513.778$ & $513.778$ \\
Total optomechanical loading & $R_{EM}^{opt}$& $90.314\;\mathrm{\Omega}$ & $90.314\;\mathrm{\Omega}$ & $90.229\;\mathrm{\Omega}$ & $90.229\;\mathrm{\Omega}$ \\
\colrule
Matching capacitance & $C_T$& $522.346 \; \textrm{fF}$ & $0 \; \textrm{fF}$  & $522.680 \; \textrm{fF}$ & $0 \; \textrm{fF}$\\
Matching inductance & $L$& $3.246 \; \mathrm{n H}$ & $3.241 \; \mathrm{nH}$ &$3.246 \; \mathrm{n H}$ & $3.239 \; \mathrm{n H}$ \\
Matching circuit frequency & $\omega_{LC}/2\pi$& $3.287 \; \mathrm{GHz}$ & $3.292\; \mathrm{GHz}$  & $3.287\; \mathrm{GHz}$ & $3.292\; \mathrm{GHz}$\\
Matching circuit quality factor & $Q_{LC}$& $1.347$ & $2.558$ & $1.347$ & $2.558$ \\
Electromechanical coupling rate & $g_{EM}/2\pi$& $56.640 \; \mathrm{MHz}$ & $107.275 \; \mathrm{MHz}$ & $56.627\; \mathrm{MHz}$ & $107.475\; \mathrm{MHz}$ \\
Electromechanical cooperativity & $C_{EM}$& $2.025$ & $7.304$ & $2.024$ & $7.301$ \\
Electromechanical loading & $R_{EM}$& $90.314\; \mathrm{\Omega}$ & $325.686\; \mathrm{\Omega}$ & $90.229\; \mathrm{\Omega}$ & $325.533\; \mathrm{\Omega}$ \\
\colrule
Conversion efficiency (standard) & $\eta$ & $42.197\;\%$ & $28.688\;\%$  & $42.158\;\%$ & $28.655\;\%$\\
Conversion efficiency (alt)& $\eta_{alt}$ & $42.197\;\%$ & $42.197\;\%$  &$14.556\;\%$ & $14.556\;\%$\\
Conversion bandwidth & $\Delta \omega /2\pi$ & $10.533\; \mathrm{MHz}$ & $24.257\; \mathrm{MHz}$  & $10.522\; \mathrm{MHz}$ &$24.243\; \mathrm{MHz}$\\
Raman noise quanta& $n_o$&  $6.598\times 10^{-5}$ & $1.829\times 10^{-5}$  & $6.591\times10^{-5}$ & $1.827\times10^{-5}$\\Thermal noise quanta& $n_{th}$& $2.203\times 10^{-2}$ & $6.110\times10^{-3}$  & $2.205\times10^{-2}$ & $6.112\times10^{-2}$\\
\end{tabular}
\end{ruledtabular}
\end{table*}

\begin{figure}
\includegraphics[width = 0.4\textwidth] {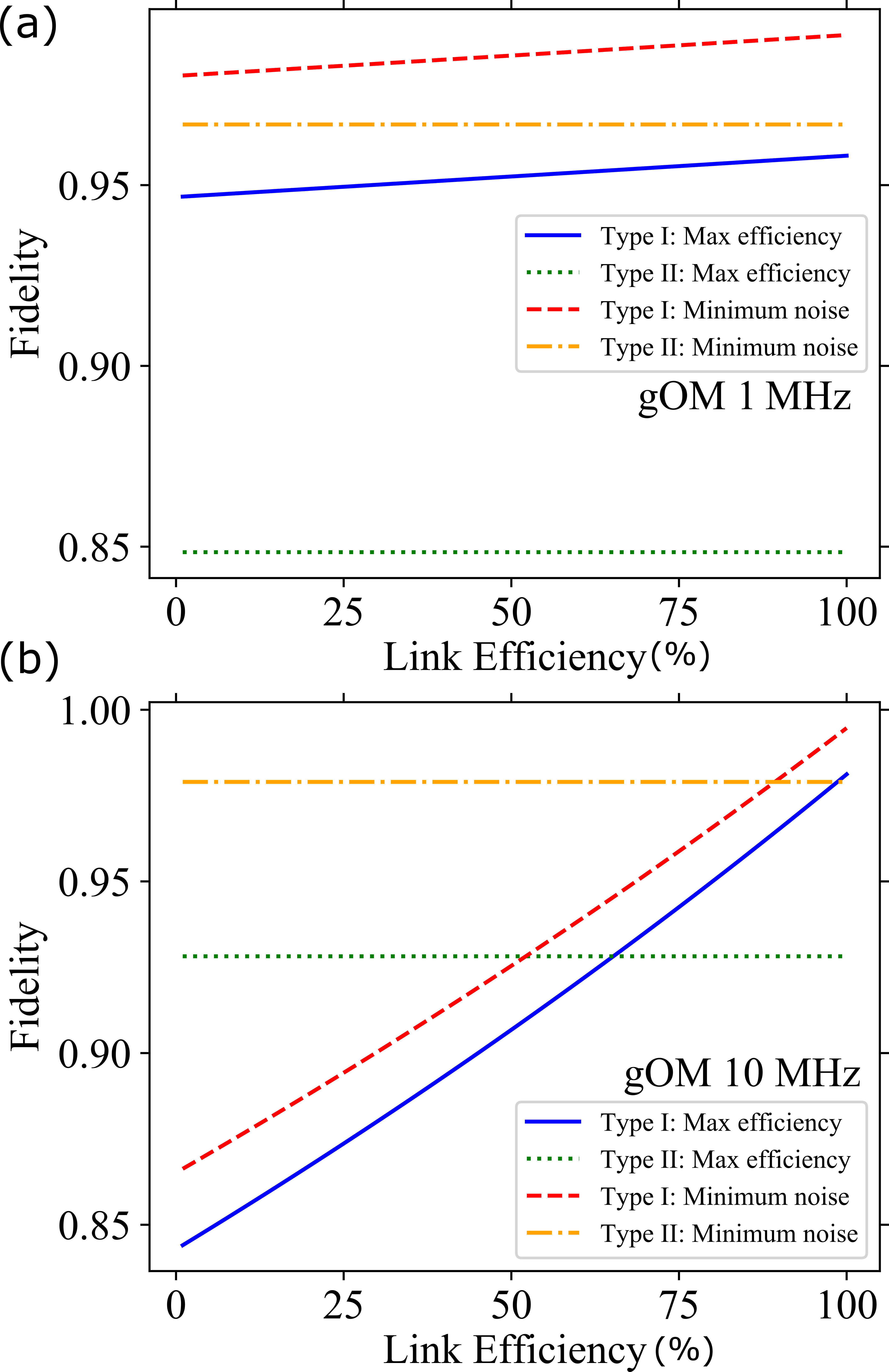}
\caption{Entanglement readout fidelity for Type-I and Type-II versus link efficiency for (a)$g_{OM}= 1\;\mathrm{MHz}$ and (b) $g_{OM}= 10\;\mathrm{MHz}$ at $T=50\;\mathrm{mK}$. Red curves: Type-I, minimize noise, Blue curves: Type-I, maximize efficiency Gold curves: Type-II, minimize noise, Green curves: Type-I, maximize efficiency.}
\label{fig:Fig8}
\end{figure}

\begin{table*}[htpb]
\caption{\label{tab:results2}%
Example results for a single-ring FBAR transducer, $T = 50\;\mathrm{mK}$, $g_{OM} = 10\;\mathrm{MHz}$ ($6.25 \times 10^8$ photons in optical cavity), $\eta_l = 50\%$. Qubit reset time is $1\;\mathrm{\mu s}$, time-bin separation $t_{sep}=\Delta_t + t_r$ and detector efficiency is $90\%$.}
\begin{ruledtabular}
\begin{tabular}{lccccr}
\textrm{Name}&
\textrm{Symbol}&
\textrm{Type I Max Eff}&
\textrm{Type I Min Noise}&
\textrm{Type II Max Eff}&
\textrm{Type II Min Noise}\\
\colrule

Transduction time & $\Delta t$ & $125.655\;\textrm{ns}$& $117.304\;\textrm{ns}$ & $125.655\;\textrm{ns}$ & $117.304\;\textrm{ns}$ \\
Photon emission rate & $r_o$ & $2.222\;\textrm{MHz}$& $2.222\;\textrm{MHz}$& $2.222\;\textrm{MHz}$& $2.222\;\textrm{MHz}$\\
Photon emission probability & $P_1$ & $24.364\%$& $22.947\% $& $24.364\%$& $22.947\%$\\
Thermal emission probability & $P_{th}$ & $0.613\%$& $0.159\%$& $0.613\%$& $0.159\%$\\
Photon generation rate & $n_r$ or $n_r'$ & $375.252\;\textrm{kHz}$& $359.542\;\textrm{kHz}$& $187.626\;\textrm{kHz}$& $179.771\;\textrm{kHz}$\\
Entanglement rate & $\tau_{ent}$ & $168.863\;\textrm{kHz}$& $161.794\;\textrm{kHz}$& $37.994\;\textrm{kHz}$& $36.403\;\textrm{kHz}$\\
Fidelity  & F & $90.678\%$& $92.541\%$& $92.819\%$& $97.898\%$\\
\end{tabular}
\end{ruledtabular}
\end{table*}

\begin{table*}[htpb]
\caption{\label{tab:results3}%
Example results for a double-ring FBAR transducer, $T = 50\;\mathrm{mK}$, $g_{OM} = 10\;\mathrm{MHz}$ ($6.25 \times 10^8$ photons in optical cavity), $\eta_l = 50\%$. Qubit reset time is $1\;\mathrm{\mu s}$, time-bin separation $t_{sep}=\Delta_t + t_r$ and detector efficiency is $90\%$.}
\begin{ruledtabular}
\begin{tabular}{lccccr}
\textrm{Name}&
\textrm{Symbol}&
\textrm{Type I Max Eff}&
\textrm{Type I Min Noise}&
\textrm{Type II Max Eff}&
\textrm{Type II Min Noise}\\
\colrule

Transduction time & $\Delta t_{2MRR}$ & $126.248\;\textrm{ns}$& $117.893\;\textrm{ns}$& $126.248\;\textrm{ns}$& $117.893\;\textrm{ns}$\\
Photon emission rate & $r_o$ & $2.222\;\textrm{MHz}$& $2.222\;\textrm{MHz}$& $2.222\;\textrm{MHz}$& $2.222\;\textrm{MHz}$\\
Photon emission probability & $P_1$ & $24.463\%$& $23.048\% $& $24.463\%$& $23.048\% $\\
Thermal emission probability & $P_{th}$ & $0.617\%$& $0.160\%$& $0.617\%$& $0.160\%$\\
Photon generation rate & n& $376.327\;\textrm{kHz}$& $360.683\;\textrm{kHz}$& $188.163\;\textrm{kHz}$& $180.341\;\textrm{kHz}$\\
Entanglement rate & $\tau_{ent}$ & $169.347\;\textrm{kHz}$& $162.307\;\textrm{kHz}$& $38.103\;\textrm{kHz}$& $36.519\;\textrm{kHz}$\\
Fidelity  & F & $90.640\%$& $92.504\%$& $92.817\%$& $97.899\%$\\
\end{tabular}
\end{ruledtabular}
\end{table*}

\begin{figure*}
\includegraphics[width = 1\textwidth] {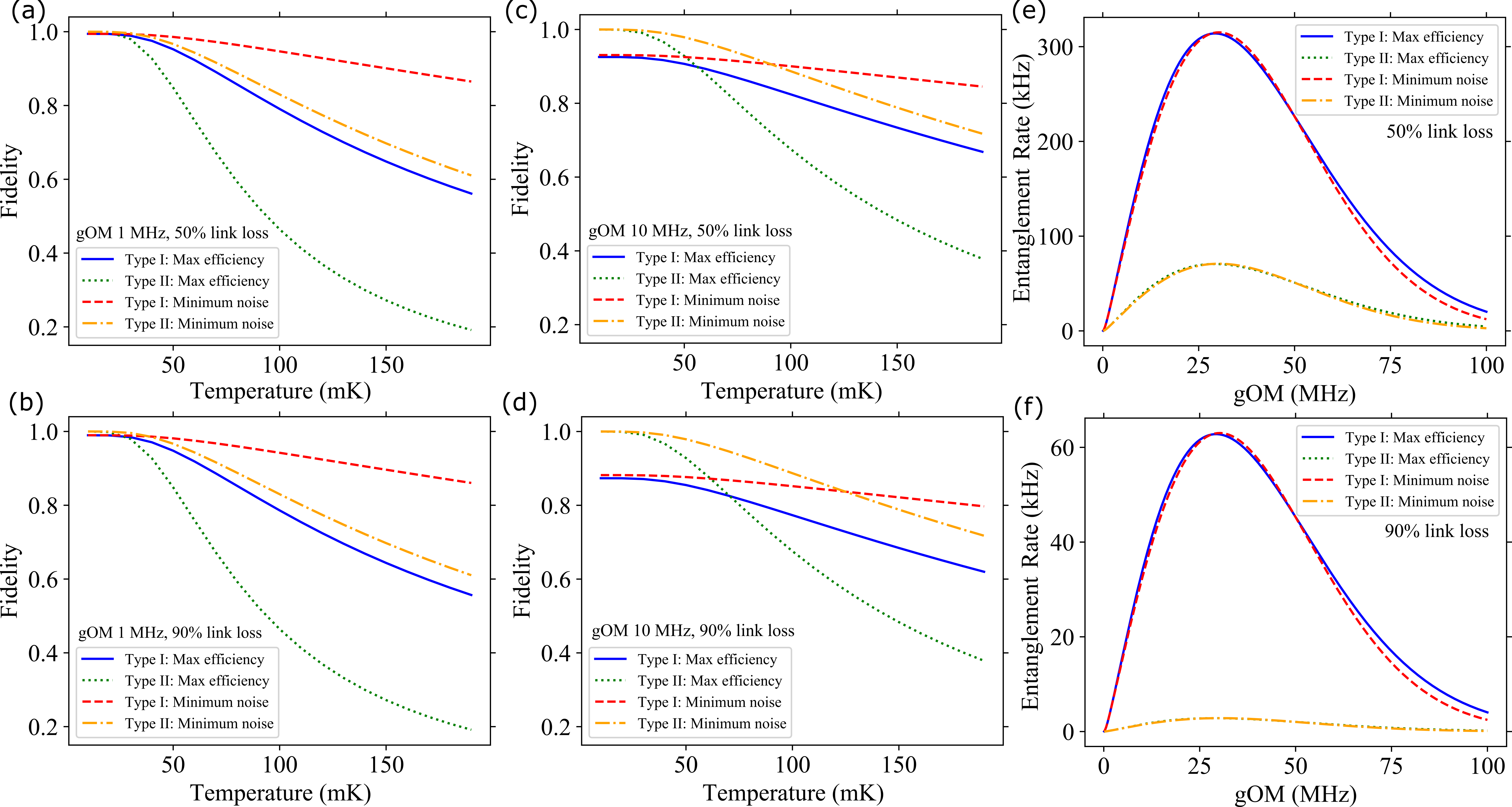}
\caption{Type I and Type-II Entanglement fidelity and rate for maximum-efficiency and minimum-noise matching networks, versus temperature, for (a)$g_{OM}= 1\;\mathrm{MHz}$ and $\eta_l = 0.5$, (b)$g_{OM}= 1\;\mathrm{MHz}$ and $\eta_l = 0.1$, (c) $g_{OM}= 10\;\mathrm{MHz}$ and $\eta_l = 0.5$ and (d) $g_{OM}= 10\;\mathrm{MHz}$ and $\eta_l = 0.1$. The entanglement fidelity degrades for the Type-I protocol at higher pump powers (higher $g_{OM}$), though using a noise-minimizing matching circuit mitigates this effect. At high temperatures, the fidelity is the worst for the efficiency-maximizing devices, regardless of whether a Type-I or Type-II protocol is used.  At low temperatures, the optimal configuration depends on $g_{OM}$. For a realistic link efficiency of $\eta_l=0.1$ (b,d), the Type-II protocol with a noise-minimizing matching circuit results in the highest entanglement fidelity at $<50\;\mathrm{mK}$. Plots (e,f) show the entanglement generation rate versus $g_{OM}$ for (e) $\eta_l = 0.5$ link loss and (f)$\eta_l = 0.1$. The $g_{OM}$ values extend out to $50\;\mathrm{MHz}$, above which the matching circuit elements become un-physical. As anticipated, Type-II protocols achieve lower entanglement rates.}
\label{fig:Fig9}
\end{figure*}

Let us now formulate expressions for the fidelity of the two-photon entanglement heralding protocol. Entanglement attempts that produce zero or one photons instead of two can be discarded: this post-selection process is powerful in that it avoids many would-be sources of infidelity that do not produce one click each in the Early and Late time bins. We are free to increase the pump power to the transducer without worrying about two-photon emission events. Following our reasoning for the fidelity expression in Equation~\ref{fid T1 no noise}, there are no sources of infidelity in the absence of noise. This will not be the case for realistic experimental conditions, but, seeing as state preparation for superconducting qubits is a high-fidelity operation, we assume as much here.

As we will discover, the presence of thermal excitations in the transducer's acoustic cavity contribute more significantly to the infidelity for the Type-II protocol, as compared to the Type-I protocol. This result may seem counter-intuitive; however it is the case, because the time-bin Type-II protocol allows more opportunities for spurious photon emission events to produce the desired click condition and generating false positive events, as detailed below.

There are two sets of false positive events to consider in the presence of thermal noise: first, for each entanglement attempt, it may happen that one node emits an entangled photon in one time bin, whereas either node emits a photon in the opposite time bin due to the presence of thermal quanta in the transducer. There are eight such contributions, two each from the four combinations of Early/Late signal photon emission: for each combination, either node may emit the noise photon. Here we note that a single node can emit in \emph{both} of the time bins as a result of thermal noise. This situation obtains when the time bin separation $t_{sep}$ is greater than the time it takes to transduce a noise phonon to an optical photon. Here we have defined $t_{sep}$ to always be greater than this value.

Second, it may happen that both clicks result from thermal emission from either node, one photon in each time bin. There are four such contributions. For the infidelity expression, we have 
\begin{multline} 
    P_{in} = 2[P_{A,E}(1-P_{B,L})P_{th}(1-P_{th}) + \\P_{A,L}(1-P_{B,E})P_{th}(1-P_{th})+\\P_{B,E}(1-P_{A,L})P_{th}(1-P_{th}) +\\ P_{B,L}(1-P_{A,E})P_{th}(1-P_{th}) +\\ (1-P_{A,E})(1-P_{B,L})P_{th}^2+\\ (1-P_{A,L})(1-P_{B,E})P_{th}^2 ]\eta_l^2\\
    =(8P_{01}P_{th}(1-P_{th})+4P_{00}P_{th}^2)\eta_l^2 
\end{multline}
where the first four terms come from signal/noise false positive events, and the final four terms come from noise/noise false positive events. So the fidelity expression takes the form
\begin{multline}
    F = \frac{2P_1^2(1-P_{th})^2}{8P_{01}P_{th}(1-P_{th})+4P_{00}P_{th}^2+ 2P_1^2(1-P_{th})^2},
 \end{multline}
where the $\eta_l^2$ terms cancel out, rendering the Type-II fidelity independent of link loss (see Figure \ref{fig:Fig9}). 

For the Type-II infidelity, do we also need to include terms corresponding to thermal photon emission \emph{in addition to} the emission of two signal photons, e.g. a three-photon emission event, where one of the emitted photons is lost before reaching the photodetectors? Fortunately, the answer is no. If we have a situation where the two-click condition is satisfied by one signal photon each in the Early and Late time bins, and we have successfully generated the distributed superconducting qubit Bell pair (Equation \ref{eq:SQ_Bell}), the presence of an additional thermal phonon within the transducer will not affect the distributed Bell state. The emissive nature of the direct transduction protocol protects us from this potential source of readout infidelity, assuming that the superconducting qubit is protected by the spurious phonon excitation via a tunable coupling element \cite{Campbell2023} placed between the transducer and the superconducting qubit that protects the qubit from spurious microwave photon absorption. This protection is not necessarily present for a blue-detuned protocol, as discussed in Appendix \ref{app:blue}.

The single-photon generation rate for Type-II protocols is, assuming $\Delta t$ and $t_r$ are the same for both nodes, the same as for Type-I protocols (Equation~\ref{eq: 1 photon gen rate}). If we use time-bin encoding, the overall entanglement generation rate is lowered by the time-bin separation $t_{sep}$, because we need two photon emission events to create our desired Bell state: one each in the Early and Late time bins. The total time it takes to get the two desired emission events is 
\begin{equation}\label{2 photon gen time T2}
    T_{2phot} = 1/(n_r') = \frac{1}{n_r}+ t_{sep}-(\Delta_t + t_r) + \frac{1}{n_r}
\end{equation}
Where $n_r'$ is the modified photon emission rate. If we set $t_{sep} = \Delta_t + t_r$, we get
 \begin{equation}\label{n photon gen time T2}
    n_r' = \frac{n_r}{2}
 \end{equation}
and the expression for the entanglement generation rate is
 \begin{equation}\label{ent gen rate T2}
    \tau_{ent} = n_r' \eta_{l}^2 \eta_{det}^2
 \end{equation}
where the link and detection efficiencies are squares to account for the fact that we need two photon detection events for the protocol to succeed. 

\subsection{\label{Results}Entanglement Generation Results}
For an FBAR transducer with the parameters summarized in Tables~\ref{tab:blesin},~\ref{tab:results}, results are plotted in Figures~\ref{fig:Fig5} ---\ref{fig:Fig9} and summarized in Table~\ref{tab:results2}. Note that all results shown are for single-ring devices, with the understanding that results will be similar for two-ring devices. 

To ascertain a numerical estimate for the fidelity of the entanglement heralding process for the FBAR transducer, we need to obtain values for $P_1$, $P_{th}$, and the associated photon generation rates. For the device parameters listed in Table~\ref{tab:blesin}, and assuming it takes 1 $\mathrm{\mu s}$ to reset the superconducting qubits between each attempt, we get the values shown in Table~\ref{tab:results2} for a single-ring device and Table \ref{tab:results3} for a double-ring device. 

For an FBAR transducer operating at $T = 50$ $\mathrm{mK}$ and $g_{OM}=10$ $\mathrm{MHz}$, the Type-I protocol has a higher entanglement generation rate: $169$ $\mathrm{kHz}$ for an efficiency-maximizing circuit and $162$ $\mathrm{kHz}$ for a noise-minimizing circuit, and a fidelity of $\sim90\%$. For a Type-II protocol, we get entanglement generation rates of about $37\;\mathrm{kHz}$. For both protocols, the fidelity is sensitive to which matching circuit is used, but to different effects. For Type-I, a fidelity of $\sim 90\%$ is achieved for efficiency maximization, and $\sim 92\%$ for noise minimization. The increase in fidelity for noise minimization can be attributed to the decrease in false detection events at the high pump power corresponding to  $g_{OM}=10$ $\mathrm{MHz}$. For Type-II, a fidelity of $\sim 93\%$ is achieved for efficiency maximization, and $\sim 98\%$ for noise minimization. Note that the entanglement generation rate is lower for noise-minimizing circuits, despite the fact that the transduction time is faster, due to the slightly lower photon emission probability.

Zeuthen \emph{et al.}~\cite{Zeuthen2020} found that, for low added noise $n_{th}<<1$ in their transducer, the two-photon protocol is less sensitive to added noise than the one-photon protocol. For the time-bin qubit-based Type-II protocol explored here, the many contributions to infidelity from the presence of thermal excitations make the protocol more sensitive to added noise. As such, the effective implementation of noise-minimizing matching circuits, with low parasitic capacitance, is paramount to high fidelity operation.

In Fig.~\ref{fig:Fig5} we plot the fidelity of a Type-I protocol versus the device temperature and optical cavity photon number. The photon numbers plotted correspond to a range $g_{OM} \in [1\;\mathrm{MHz}, 10\;\mathrm{MHz}]$. The top row of the figure shows the fidelity for transducers with efficiency-maximizing circuits, whereas the bottom row has noise-minimizing circuits. Overall, using a noise-minimizing matching circuit maximizes fidelity. The fidelity difference between the two options is more pronounced for lower $\eta_{l}$ values. The true value of the noise-minimizing circuits becomes clear at higher temperatures. For example, with $\eta_{l} = 50\%$, at $150 \;\mathrm{mK}$ the entanglement fidelity is $80-90\%$ with a noise-minimizing circuit, but is only $60-70\%$ for an efficiency-maximizing circuit. 

In Figure~\ref{fig:Fig9} we plot the fidelity of Type-I and Type-II protocols vs. temperature for efficiency-maximizing and noise-minimizing matching circuits. These are plotted for different $g_{OM}$ values and link efficiencies. For realistic link losses of $90\%$, we see that Type-II protocols result in higher readout fidelities at higher pump powers, but the Type-II fidelity falls off more steeply with increasing temperature, as these are more sensitive to thermal noise. For a lower $g_{OM}$ in panel (b), the Type-I protocol for a device with a noise-minimizing matching circuit becomes the highest-fidelity option for temperatures of $50 \;\mathrm{mK}$ and above. 

\begin{figure}
\includegraphics[width = 0.5\textwidth]{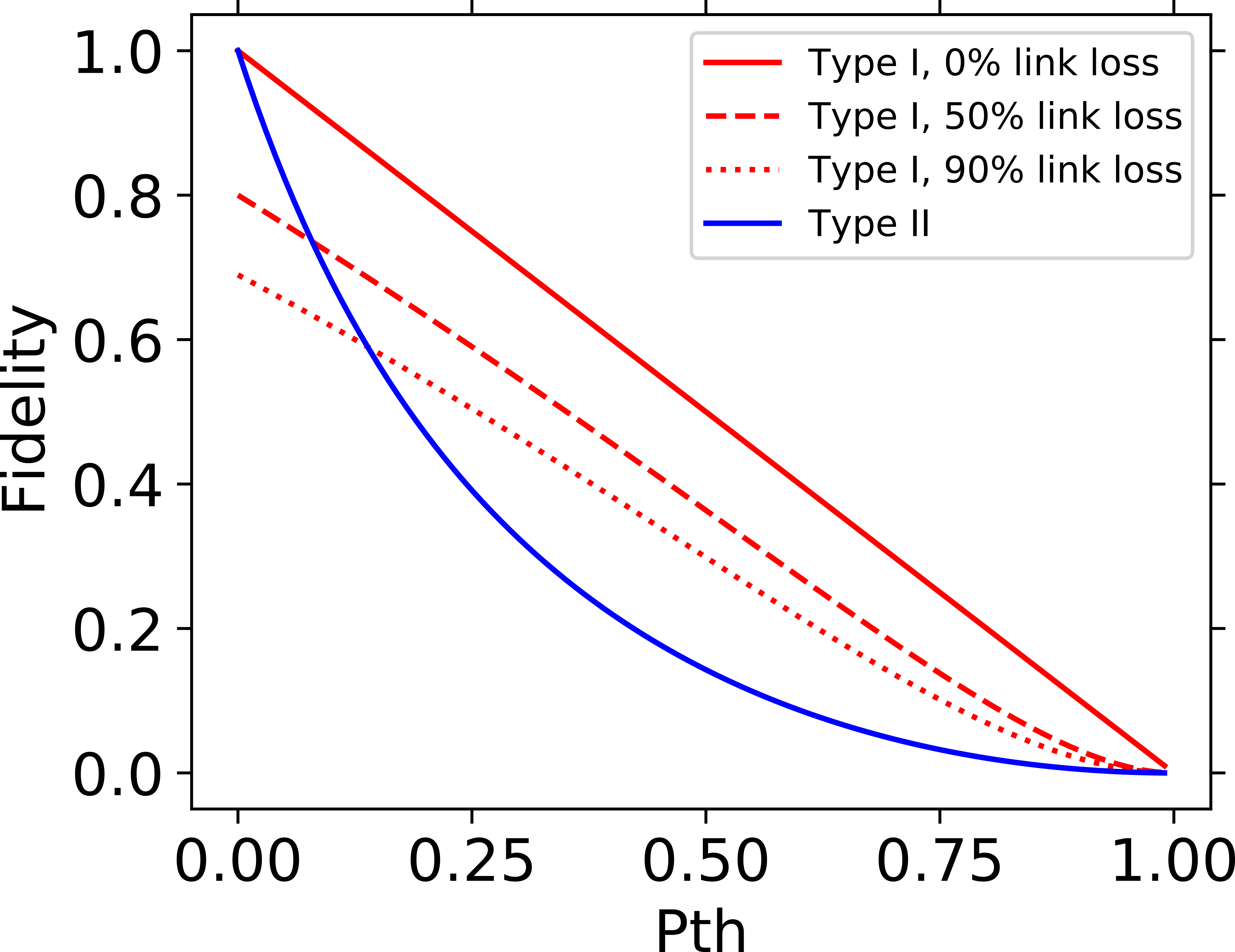}
\caption{A comparison of the dependence of Type-I and Type-II readout fidelity on $P_{th}$. While the Type-II fidelity does not depend on $\eta_{l}$, it falls off more rapidly with increasing $P_{th}$.}
\label{fig:Fig10}
\end{figure}

To better understand the relative dependence of the Type-I and Type-II protocols' fidelity on thermal noise, we write out simplified expressions for each protocol's dependence on $P_{th}$: 
\begin{multline}
    F_{Type  I} \sim \\ \frac{2(1-P_{th})^2}{(1-P_{th})^2(3-\eta_l)+P_{th}(1-P_{th})(3-\eta_l)+P_{th}^2(1-\eta_l)} \\ \xrightarrow[P_{th} \rightarrow 0]{} \frac{2}{3 - \eta_l}\\
    F_{Type II} \sim \frac{-(1-P_{th})^2}{P_{th}^2 - 2P_{th}- 1}  \xrightarrow[P_{th} \rightarrow 0]{} 1\\
\end{multline}
These expressions are plotted versus $P_{th}$ in Figure \ref{fig:Fig10}, showing how the Type-II readout fidelity falls off steeply with increasing $P_{th}$. This sensitive dependence on noise further motivates the implementation of noise-minimizing impedance matching networks for the FBAR transducer.

Next, we compare the entanglement generation rate between the two protocols. The results show that higher entanglement generation rates are achievable with Type-I protocols, which is to be expected. For Type-I, using a noise-minimizing circuit decreases the entanglement generation rate by $\sim5\%$ while achieving a slightly improved readout fidelity by $\sim2\%$. As such, the use of efficiency-maximizing circuits will be preferable at low temperatures below $\sim75\;\mathrm{mK}$, since they will achieve a higher entanglement rate at about the same fidelity and will be easier to fabricate. For the Type-II protocol, doing the same decreases the entanglement generation rate by  $\sim5\%$, while boosting fidelity from $92\%$ to $97\%$; given the relatively small decrease in entanglement rate, the boost in fidelity justifies the prioritization of noise-minimizing matching circuits, despite the inevitable difficulties in their fabrication.

\section{\label{Discussion}Discussion and Outlook}
As summarized in Table \ref{tab:results2}, for superconducting qubits networked using impedance-matched FBAR transducers, in a realistic operation regime, we show that entanglement can be achieved at rates of up to $168\;\mathrm{kHz}$ and fidelities of $~90-98\%$,  at a temperature of $50\;\mathrm{mK}$, depending on the matching circuit and the protocol in use. If the superconducting qubits have coherence times of $100 \;\mathrm{\mu s}$, or equivalently, decoherence rates of $10 \;\mathrm{kHz}$, we can in principle achieve entanglement at a rate exceeding the decoherence rate of the qubits. With ever-increasing qubit coherence times \cite{Wang2022t}, it will only become easier to reach this performance threshold. This work only explored impedance matching circuits for one set of physical parameters for the FBAR transducer. In practice, the transducer's internal properties, such as its constituent materials, their geometry, and the external optical coupling rate $\kappa_{ext}$ can be modified to further improve performance tailored for a specific application.

Unlike OMC-based optomechanical transducers that exhibit single-photon optomechanical coupling rates of $10-100\;\mathrm{kHz}$ or more, FBAR transducers will have values of $<1 \;\mathrm{kHz}$. While the lower coupling rate can be viewed as a disadvantage, our results show that it can be beneficial. The comparatively low optomechanical loading $R^{opt}_{EM}$ for the FBAR transducers at a given optical pump power requires larger matching capacitances and lower matching inductances, as compared to OMC-based transducers \cite{Wu2020},  which makes the design and fabrication of the matching networks more feasible. This allows the transducer's piezo-mechanical components to have a higher static capacitance, enabling realistic fabrication targets, and allows for more accurate matching circuit designs factoring in non-idealities such as non-zero resistance $R_L$ and parasitic capacitances and inductances.

Going forward, our model can be improved by treating the optomechanical coupling rate and temperature of the device as dependent variables, instead of independent variables. To do this, a mathematical relationship between the optical pump power to the device and the device's temperature must be established. The relationship will depend on the properties of the transducer's optical cavity, the facet loss and scattering at the optical fiber-to-chip interface, the degree of thermal anchoring of the device packaging to the cryostat, the cryostat's cooling power, and more. Defining the relationship between $g_{OM}$ and device temperature for FBAR transducers is an active area of experimental and theoretical exploration. Results obtained from simulations \cite{Ivira2008} and measurements in the near term will be used to clarify this relationship. 

\begin{acknowledgments}
We wish to thank Hao Tian, Alaina Attanasio, and Sunil Bhave of the OxideMEMS laboratory at Purdue University for providing our team with FBAR transducers, helping us learn about how they work, and for a fruitful collaboration.

\end{acknowledgments}

\section{Author contributions}
E.S., S.S., M.S., D.C. and T.W. designed transducers and matching networks and performed simulations to verify design parameters. E.S., Z.S., D.H. and D.C. designed the entanglement heralding protocols. E.S., E.A., J.S., S.S. and S.M. performed theoretical calculations. E.S. and M.L. directed and supervised the project. All authors contributed to writing the manuscript.
\section{Conflicts of interest}
The authors declare no conflicts of interest. 

\section{Funding acknowledgments}
E.S. gratefully acknowledges funding from the Air Force Office of Scientific Research's Protostar LRIR program, under award 25RICOR003. 

\appendix

\section{\label{app:symbols}Comprehensive symbol table}
\begin{table*}[htpb]
\caption{List of Symbols and Definitions}
\label{tab:symbols}
\begin{ruledtabular}
\begin{tabular}{lcl}
\textrm{Symbol} & \textrm{Units} & \textrm{Definition} \\
\colrule
\multicolumn{3}{c}{\textbf{Mechanical Parameters}} \\
\colrule
$\omega_m$ & rad/s & Mechanical mode frequency \\
$\omega_s$ & rad/s & Series resonant frequency \\
$\gamma_m$ & rad/s & Intrinsic mechanical linewidth \\
$C_m$ & F & Motional capacitance \\
$L_m$ & H & Motional inductance \\
$R_m$ & $\Omega$ & Motional resistance \\
$k_{eff}^2$ & - & Electromechanical coupling factor \\
$g_{EM}$ & rad/s & Electromechanical coupling rate \\
$\mathcal{C}_{EM}$ & - & Electromechanical cooperativity \\
\colrule
\multicolumn{3}{c}{\textbf{Optical Parameters}} \\
\colrule
$\omega_o$ & rad/s & Optical cavity frequency \\
$\kappa_o$ & rad/s & Total optical linewidth \\
$\kappa_i$ & rad/s & Intrinsic optical linewidth \\
$\kappa_{ext}$ & rad/s & External optical coupling rate \\
$g_{OM}$ & rad/s & Optomechanical coupling rate \\
$\mathcal{C}_{OM}$ & - & Optomechanical cooperativity \\
$\mathcal{C}_{OO}$ & - & Optical-optical cooperativity \\
$J$ & rad/s & Ring-ring coupling rate \\
$\eta_o$ & - & Optical external coupling efficiency \\
$\mathscr{L}_{\pm}^2$ & - & Lorentzian sideband amplitudes \\
$\Delta$ & rad/s & Optical pump detuning \\
\colrule
\multicolumn{3}{c}{\textbf{Electrical/Matching Circuit}} \\
\colrule
$C_0$ & F & Static capacitance \\
$R_0$ & $\Omega$ & Static resistance \\
$C_T$ & F & Matching capacitance \\
$L_T$ & H & Matching inductance \\
$R_L$ & $\Omega$ & Matching inductor resistance \\
$Z_{tx}$ & $\Omega$ & Transmission line impedance \\
$\omega_{LC}$ & rad/s & Matching circuit resonant frequency \\
$Q_{LC}$ & - & Matching circuit quality factor \\
$\kappa_e$ & rad/s & Electrical linewidth \\
$\eta_e$ & - & Electrical external coupling efficiency \\
\colrule
\multicolumn{3}{c}{\textbf{Equivalent Circuit Elements}} \\
\colrule
$R_{EM}$ & $\Omega$ & Electromechanical loading resistance \\
$R_{EM}^{opt}$ & $\Omega$ & Optimal electromechanical loading \\
$R_{OM}$ & $\Omega$ & Optomechanical coupling resistance \\
$R_{OM}^{eq}$ & $\Omega$ & Equivalent optomechanical resistance \\
$R_{OM,\pm}$ & $\Omega$ & Upper/lower sideband resistances \\
$R_{OO,\pm}$ & $\Omega$ & Optical-optical sideband resistances \\
\colrule
\multicolumn{3}{c}{\textbf{Performance Metrics}} \\
\colrule
$\eta$ & - & Conversion efficiency \\
$\eta_{alt}$ & - & Alternative efficiency expression \\
$n_o$ & - & Raman noise quanta per photon \\
$n_{th}$ & - & Thermal noise quanta per photon \\
$n_m$ & - & Thermal bath occupancy \\
$\Delta \omega$ & rad/s & Conversion bandwidth \\
\colrule
\multicolumn{3}{c}{\textbf{Entanglement Protocol Parameters}} \\
\colrule
$\Delta t$ & s & Transduction time duration \\
$r_0$ & Hz & Photon generation rate \\
$P_1$ & - & Single photon emission probability \\
$P_{th}$ & - & Thermal photon emission probability \\
$P_{01}, P_{10}$ & - & Target Bell state probabilities \\
$F$ & - & Readout fidelity \\
$\tau_{ent}$ & Hz & Entanglement generation rate \\
$\eta_l$ & - & Link efficiency \\
$\eta_{det}$ & - & Detector efficiency \\
$t_r$ & s & Qubit reset time \\
\end{tabular}
\end{ruledtabular}
\end{table*}

\section{\label{app:staticR}Adding transducer static resistance into Extended BVD model}

As mentioned in the main text, the mBVD model captures the behavior of piezo-mechanical devices more accurately than the standard BVD model. As such, here, we re-derive relevant expressions for a transducer's equivalent circuit with a nonzero static resistance $R_0$ within the piezo-mechanical sub-circuit.

The electrical LC impedance becomes
\begin{multline}
    Z_e(\omega) = -i\omega L +Z_{tx} + R_L + \frac{1}{-i\omega(C_0+C_T) } \rightarrow \\
    Z_e'(\omega) = -i\omega L + \left( \frac{1}{Z_{tx}+R_L} + \frac{1}{R_0} \right)^{-1} + \frac{1}{-i\omega(C_0+C_T) }
\end{multline}

This new impedance leads to a modified quality factor of the LC transformer:
\begin{equation}\label{qualityprime}
    Q_{LC}' = \frac{1}{\left[(Z_{tx}+R_L)^{-1} + R_0^{-1}\right]^{-1}} \sqrt{\frac{L}{C_T + C_0}}
\end{equation}
with an associated electromechanical loading factor
\begin{equation}
    R_{EM}' =Q_{LC}'^2\left[(Z_{tx}+R_L)^{-1} + R_0^{-1}\right]^{-1}. 
\end{equation}

Matching circuit expressions, including the matching capacitance
\begin{equation}
    C_T' = \frac{C_m}{2}\left[ \sqrt{1+\frac{4}{R_{EM}^{opt} \omega_s^2 C_m^2\left( \frac{(Z_{tx}+R_L)R_0}{Z_{tx}+R_L+R_0}\right)}}-1\right] -C_0
\end{equation}
angular frequency
\begin{equation}
\omega_m' = \sqrt{\frac{1}{L_m}\left(\frac{1}{C_m}+\frac{1}{C_T'+C_0}\right)}
\end{equation}
and inductance
\begin{equation}
    L' = \frac{1}{\omega_m'}\sqrt{R_{EM}^{opt}\left( \frac{(Z_{tx}+R_L)R_0}{Z_{tx}+R_L+R_0}\right)}.
\end{equation}

\section{\label{app:losss}\texorpdfstring{$R_L \neq 0\;$} from dielectric loss}

In the main text, as in previous work, we assumed that for a superconducting circuit, the matching circuit is lossless, leading to a resistance $R_L = 0\;\mathrm{\Omega}$ in the equivalent circuit. For real devices, there will be nonzero loss. 

Here we consider a matching network that is fabricated on a substrate with a nonzero loss tangent $\tan(\delta)$. 

The matching capacitor $C_T$ will have a loss term associated with dielectric loss, referred to as an equivalent series resistance $R_{CT}$. For a dielectric layer with loss tangent $\tan(\delta)$, a capacitor with ideal capacitance $C_T$ at angular frequency $\omega_{LC}$ has \cite{Riad2020} 
\begin{equation}
    R_{CT} = \frac{\tan(\delta)}{\omega_{LC} C_T}
\end{equation}

Similarly, the matching inductor $L$ has an equivalent resistance
\begin{equation}
    R_{ind} = \tan(\delta)\omega_{LC} L.
\end{equation}
The quality factor of the matching network for $R_L, R_0 \neq 0$ is shown in Equation~\ref{qualityprime}, where we set $R_L = R_{CT}+R_{ind}$.

If the matching network is comprised of an Al thin film deposited on thermal SiO\textsubscript{2} with a loss tangent of approximately $3 \times 10^{-4}$ \cite{oconnell2008}, we have
\begin{multline}
    R_{L} = tan(\delta)\left(\frac{1}{\omega_{LC}C_T}+\omega_{LC}L \right)\\
    = 47.925\;\mathrm{m\Omega}.
\end{multline}

\begin{table}[b]
\caption{\label{tab:RL}%
Difference in matching circuit (maximize efficiency) and transducer (single ring) values for $R_L=0\;\mathrm{\Omega}$ vs. $R_L \neq 0 \;\mathrm{\Omega}$.}
\begin{ruledtabular}
\begin{tabular}{lcr}
\textrm{Symbol}&
\textrm{$R_L = 0\;\mathrm{\Omega}$}&
\textrm{$R_L = 47.925\;\mathrm{m\Omega}$}\\
\colrule
 $C_m$& $0.860 \; \textrm{fF}$ & $0.860 \; \textrm{fF}$  \\
$L_m$& $2.729 \; \mathrm{\mu H}$ & $2.729 \; \mathrm{\mu H}$ \\
$R_m$& $44.588 \; \mathrm{\Omega}$ & $44.588 \; \mathrm{\Omega}$ \\
\colrule
$\mathscr{L}_+^2$& $1$ & $1$ \\
$\mathscr{L}_-^2$& $1.303\times10^{-4}$ & $1.303\times 10^{-4}$ \\
$R_{OM,+}$& $45.732\;\mathrm{\Omega}$ & $45.732\;\mathrm{\Omega}$  \\
$R_{OM,-}$& $5.959\times 10^{-3}\;\mathrm{\Omega}$ & $5.959\times 10^{-3}\;\mathrm{\Omega}$ \\
$C_{OM}$ & $1.026$ & $1.026$  \\
$R_{EM}^{opt}$& $90.314\;\mathrm{\Omega}$ & $90.314\;\mathrm{\Omega}$ \\
\colrule
$C_T$& $522.346 \; \textrm{fF}$ & $522.002 \; \textrm{fF}$  \\
$L$& $3.246 \; \mathrm{n H}$ & $3.247 \; \mathrm{nH}$  \\
$\omega_{LC}/2\pi$& $3.287 \; \mathrm{GHz}$ & $3.287\; \mathrm{GHz}$\\
$Q_{LC}$& $1.347$ & $1.347$  \\
$g_{EM}/2\pi$& $56.640 \; \mathrm{MHz}$ & $56.654 \; \mathrm{MHz}$  \\
$C_{EM}$& $2.025$ & $2.026$ \\
$R_{EM}$& $90.314\; \mathrm{\Omega}$ & $90.314\; \mathrm{\Omega}$  \\
\colrule
$\eta$ & $42.197\;\%$ & $42.197\;\%$ \\
$\eta_{alt}$ & $42.197\;\%$ & $42.197\;\%$ \\
$\Delta \omega /2\pi$ & $10.533\; \mathrm{MHz}$ & $10.533\; \mathrm{MHz}$  \\
$n_o$&  $6.598\times10^{-5}$ & $6.598\times10^{-5}$ \\
$n_{th}$& $2.203\times10^{-2}$ & $2.203\times10^{-2}$  \\
\end{tabular}
\end{ruledtabular}
\end{table}

\section{Matching inductor parasitic capacitance}\label{app:parasitic}
In this section we examine the stray capacitances and inductances of the matching elements and their effect on impedance-matched transducer design.

We start with the matching inductor. There will be two parasitic capacitive terms: the inductor's self-capacitance, and its capacitance to ground. The first arises from the inherent capacitance that exists within the inductor, including capacitances between adjacent wires within a meandering or spiral structure, and the second arises from an unwanted capacitance between the inductor metal and the nearby ground-plane metal layer. For superconducting spiral inductors, previous studies have found inductors fabricated on Silicon with inductances of $3-7\;\mathrm{nH}$ to have self-capacitances of about $5\;\mathrm{fF}$ \cite{Peruzzo2020}. Additional capacitance to the ground plane must also be considered, which may add an additional $5-10\;\mathrm{fF}$ or more \cite{Hooker2016, Xue2007}. For a nanowire inductor made of a high-kinetic inductance material such as niobium nitride (NbN) \cite{Niepce2019}, a $40\;\mathrm{nm}$-wide nanowire has an inductance of  $2.05\;\mathrm{mH/m}$ and a capacitance of $44\;\mathrm{pF/m}$. A target inductance of $3.247\;\mathrm{nH}$ would require a length of $1.584\;\mathrm{\mu m}$, corresponding to a  capacitance of $70\;\mathrm{aF}$ and a self-resonance at $2\;\mathrm{THz}$.

These considerations motivate a modification of the matching network design formalism to account for stray capacitances. The inductor's parasitic capacitance $C_p = C_s+ C_g$, where $C_s$ is its self capacitance and $C_g$ is its capacitance to ground. $C_p$ will be added in parallel to the intentionally designed matching capacitor $C_T$, leading to a modified equivalent matching capacitance $C_T'$:
\begin{equation}
    C_T' = C_T + C_p.
\end{equation}

Next we consider the stray inductance $L_{CT}$ of the matching capacitor. This parasitic term will add in series with the existing inductance $L$, for a modified equivalent matching inductance $L'$:
\begin{equation}
    L' = L + L_{CT}.
\end{equation}
A modified equivalent circuit incorporating the parasitic terms into the matching network is shown in Fig.~\ref{fig:Fig11}. To design a matching circuit that incorporates the parasitic terms, we follow the process outlined in the main text, using Equations~\ref{eq:CT general}-\ref{eq:CT closed}. The values we obtain from those expressions are $C_T'$ and $L'$. Next, we use simulations or numerical estimates to obtain values for the parasitic terms $C_p$ and $L_{CT}$. Finally, we modify the matching network components $L$, $C_T$ as follows:
\begin{multline}
    C_T = C_T' - C_p \; ; \;
    L = L' - L_{CT}.
\end{multline}
Clearly, the modified components will have slightly different parasitic values than originally estimated. If desired, this process can be iterated to obtain more and more accuracy. 

\section{Bell-State Analysis for Time-Bin Photons}
In the main text, we implement a Type-II entanglement heralding protocol based on time-bin encoding of photon qubits. The use of the time-bin basis leads to a question of how Bell state analysis will be performed. 

For time-bin photons, we have two options. We could use a balanced Bell state analyzer, within which the two path lengths are equal, or an unbalanced Bell state analyzer, in which the path length difference between the two arms is equal to the time bin separation. Let's explore the benefits and drawbacks of these two options.

Time bin qubits can be measured using a simple (balanced) Bell state analyzer with a non-polarizing 50:50 beamsplitter and two SNSPDs. With sufficient temporal resolution, this setup can be used to project a two-photon state onto $\ket{\Psi^-}$ or $\ket{\Psi^+}$. Here, the different Bell state projections correspond to different temporal detection patterns\cite{Valivarth14}. Specifically, if one detector registers a photon in the Early bin and the other detector registers a photon in the Late bin, the photons are projected into $\ket{\Psi^-}$. The $\ket{\Psi^+}$ state is projected when the same detector registers both Early and Late photons. 
The photonic Bell states, with early (E) and late (L) photons emerging from nodes A and B, take the form
\begin{equation}
\begin{gathered}
\ket{\Psi^-_{phot-A,B}} = \frac{1}{\sqrt{2}}(\ket{1_{E}0_{L}}_A\ket{0_{E}1_{L}}_B - \ket{0_{E}1_{L}}_A\ket{1_{E}0_{L}}_B)\\
\ket{\Psi^+_{phot-A,B}} = \frac{1}{\sqrt{2}}(\ket{1_{E}0_{L}}_A\ket{0_{E}1_{L}}_B + \ket{0_{E}1_{L}}_A\ket{1_{E}0_{L}}_B)
\end{gathered}
\end{equation}

If we assume that the single photon detectors employed within the BSA can resolve the arrival time of photons with a given time-bin separation, we can proceed to use a balanced interferometer for Bell state measurements. Of course, this requires the dead time of the detector to be shorter than the time bin separation. For our chosen time-bin separation of $500\;\mathrm{ns}$, this requirement can be satisfied, as commercial SNSPDs have dead times falling in the $10-100\;\mathrm{ns}$ range. Assuming both $\ket{\Psi^-}$ and $\ket{\Psi^+}$ can be resolved, the BSA has a maximum efficiency of $50 \%$. 

Alternatively, we may use an unbalanced BSA, or a time-domain interferometer (TDI). One arm of the TDI has a delay line that applies a time delay equal to the time-bin separation, $\Delta t$, allowing for the erasure of which-time-bin information. Coincidence events occur when Early photons propagate through the long arm and Late photons propagate through the short arm. This will occur $50\%$ of the time for early-late input pairs. The other $50\%$ of the time, the Late photons will travel through the long arm, and vice versa, doubling the time delay between Early and Late. If we throw away non-coincident detection events, we are only able to distinguish the $\ket{\Psi^-}$ state. 

Another drawback of this measurement method is the need to actively stabilize the interferometer’s long delay line. Fiber-based delay lines are susceptible to perturbations due to environmental vibrations and temperature fluctuations. The time delay must be stabilized using active components such as fiber piezo-stretchers controlled by feedback loops within periodic calibration measurements. This benefit of this additional overhead carries the advantage of eliminating strict requirements on the reset time of the single photon detectors  and simplifying post-detection analysis. 

\section{\label{app:infidelity}Building the Type-I conditional readout infidelity expression}

\begin{figure}
\includegraphics[width = 0.5\textwidth]{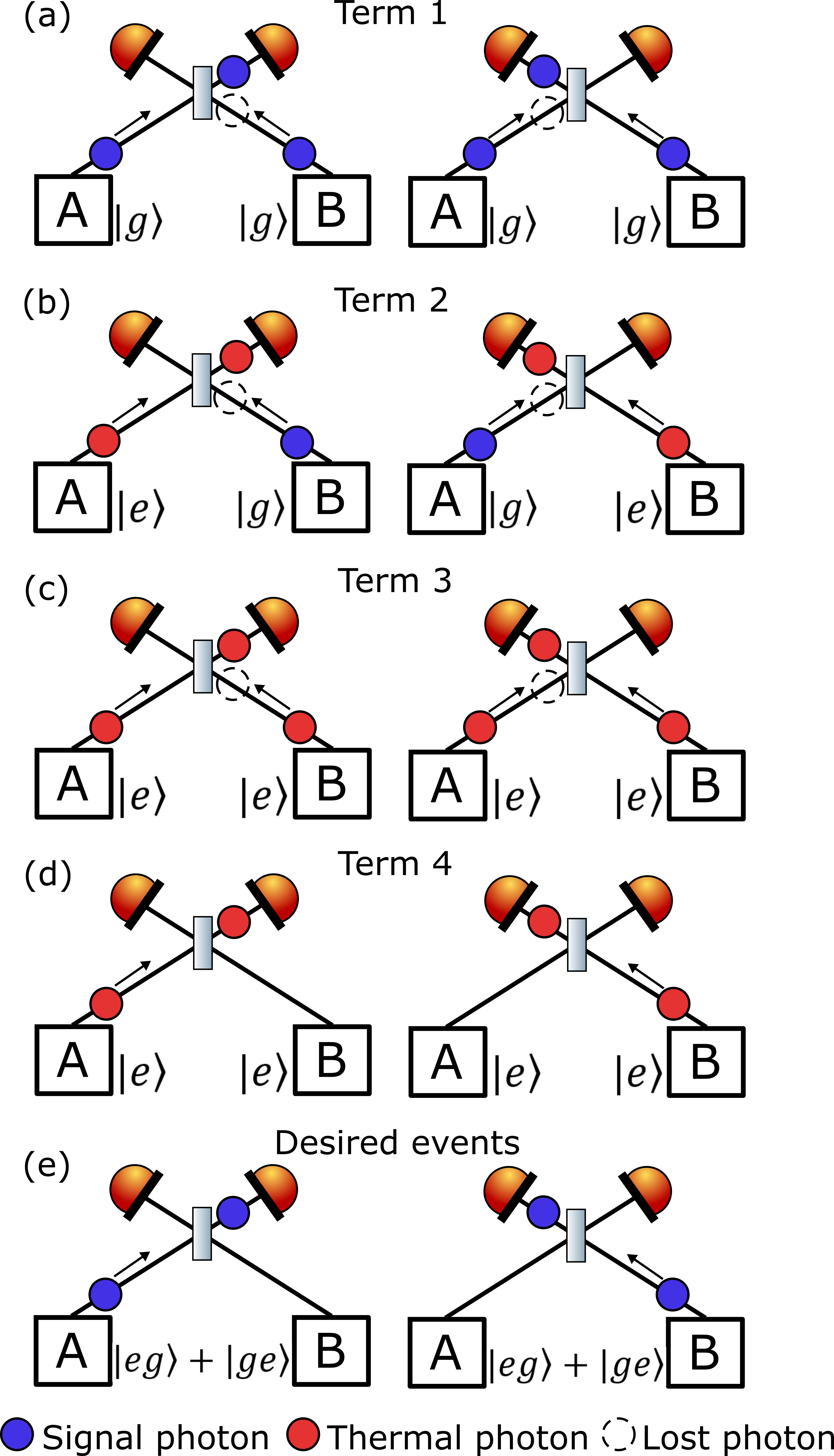}
\caption{Diagrams of scenarios contributing (a-d) to Terms 1-4 of the infidelity expression and (e) desired entanglement heralding events.}
\label{fig:Fig13}
\end{figure}

In this section, we provide a detailed physical picture underlying the structure of the fidelity expressions given in Equations \ref{typeI_infidelity} and \ref{typeI_infidelity_s}. An accompanying diagram for each scenario is shown in Fig. \ref{fig:Fig13}.

For the Type-I protocol, entanglement between remote nodes is established when one photon is detected at the Bell state analyzer. False positive events occur when one photon is detected, as desired, but the remote superconducting qubits are \emph{not} in the desired Bell state. Let us review our notation:

\begin{table}[h]
\centering
\caption{Notation for Type I Protocol Analysis}
\begin{tabular}{llc}
\hline
\textbf{Sym} & \textbf{Description} & \textbf{Expression} \\
\hline
$P_1$ & Prob. a node emits a signal photon & $1 - e^{-r_0\Delta t}$ \\
$P_0$ & Prob. a node emits no signal photon & $e^{-r_0\Delta t}$ \\
$P_{\text{th}}$ & Prob. a node emits a thermal photon & $r_0 n_{\text{th}} \Delta t$ \\
$P_{01}$ & Prob. Node A emits 1, Node B emits 0 & $P_1 P_0$ \\
$P_{10}$ & Prob. Node A emits 0, Node B emits 1 & $P_0 P_1$ \\
$P_{00}$ & Prob. both nodes emit 0 photons & $P_0^2$ \\
$\eta_l$ & Link transmission prob. & --- \\
\hline
\end{tabular}
\label{tab:type1_notation}
\end{table}
Now we will break Equation \ref{typeI_infidelity} down term by term. 
\\Term 1: $P1^2(1-P_{th})^2\eta_l(1-\eta_l)$
\\Physical process: Both nodes emit signal photons (probability $P_1^2$), neither node emits a thermal photon (probability $(1-P_{th}^2)$). One signal photon reaches a detector ($\eta_l$), and the other is lost before reaching a detector ($(1-\eta_l)$).
\\Why it's a false positive: The Bell State Analyzer sees one detector click, but both superconducting qubits are in their ground state, not in the desired $\ket{ge}\pm\ket{eg}$ Bell state.
\\
\\Term 2: $2P_{01}P_{th}(1-P_{th})\eta_l (1-\eta_l)$
\\Physical process: One node emits a signal photon (probability $P_{01}$) and no thermal photon (probability $(1-P_{th})$), and the other node emits a thermal photon (probability $P_{th}$) but no signal photon. The thermal photon reaches a detector ($\eta_l$) but the signal photon is lost before reaching a detector ($(1-\eta_l)$). A factor of two (assuming $P_{10}=P_{01}$) accounts for either node being the thermal emitter.
\\Why it's a false positive: The Bell State Analyzer sees one detector click from the thermal photon but not the signal photon. From the detector click we make the correct inference that one signal photon was emitted, but the detected photon is uncorrelated with any superconducting qubit state. Without this correlation, the qubits remain in a separable product state rather than the desired Bell state. While the qubits are in orthogonal states, detecting a thermal photon fails to erase the "which-path" information, which is needed to create a coherent Bell state between the remote qubits.
\\
\\Term 3: $P_{00}P_{th}^2\eta_l(1-\eta_l)$
\\Physical process: Neither node emits a signal photons (probability $P_{00}$). Both nodes emit thermal photons (probability $P_{th}$). One of the thermal photons reaches a photodetector ($\eta_l$), but the other is lost before reaching a detector ($(1-\eta_l)$).
\\Why it's a false positive: The Bell State Analyzer sees one detector click from thermal emission, but both superconducting qubits are in their excited state, not in the desired $\ket{ge}+\ket{eg}$ Bell state.
\\
\\Term 4: $2P_{00}P_{th}(1-P_{th})\eta_l$
\\Physical process: Neither node emits a signal photon (probability $P_{00}$). One node emits a thermal photon (probability $P_{th}$), which reaches a photodetector ($\eta_l$). A factor of two accounts for either node being the thermal emitter. 
\\Why it's a false positive: The Bell State Analyzer sees one detector click from thermal emission, but both superconducting qubits are in their excited state, not in the desired $\ket{ge}+\ket{eg}$ Bell state.
\\
Desired events: $2P_{01}(1-P_{th})^2\eta_l$ 
\\Physical process:
\\The overall conditional readout fidelity takes the form 
\begin{equation}
   F= \frac{\text{(Desired events)}}{(\text{Desired events } + \text{ False positive events})}
\end{equation}

\section{\label{app:blue}Blue-Detuned Transducer Operation}

For completeness, here we consider entanglement heralding protocols that utilize spontaneous parametric down conversion (SPDC) in a blue-detuned transducer to generate entangled microwave-telecom photon pairs. This scenario is briefly explored using a formalism similar to our own in references \cite{Krastanov2021, Wang2022} but not in \cite{Zeuthen2018}. Blue-detuned operation with a Type-II protocol using time-bin photons, but not with a Type-I protocol, is explored in depth in reference \cite{Zhong2020a}. 

For a Type-I protocol utilizing SPDC seeded by a blue-detuned transducer pump, each of two identical nodes produces entangled microwave-optical photon pairs. The microwave photon is converted to a superconducting qubit excitation via stimulated Raman absorption, while the optical photon is emitted by the transducer toward the BSA. The detection of an optical photon in the BSA (a single-click event) heralds the production of a single microwave photon. For a qubit initialized in its ground state, detection of an optical photon corresponds to the qubit being in its excited state. In a simplified picture, the qubit-cavity system begins in the state $\ket{1_{m}}\ket{g}$, where $m$ denotes the microwave mode. Absorption of the photon via a Jaynes-Cummings interaction evolves the system to the state $\ket{0_{m}}\ket{e}$. 

A Type-I heralding protocol based on this process pumps two copies of a system and, as detailed above, erases which-path information on a beamsplitter to herald a distributed microwave photon Bell pair. A successful attempt produces the target photonic Bell state in Equation \ref{eq:target_photonic}, which projects the remote superconducting qubits into the distributed Bell state
\begin{equation}
\ket{\Psi} = \sqrt{P_{01}} \ket{g_Ae_B} \pm \sqrt{P_{10}} \ket{e_Ag_B}
\end{equation}

As discussed in \cite{Krastanov2021, Zhong2020}, this protocol is susceptible to additional source of readout infidelity. SPDC-based protocols allow for multi-photon generation events \cite{Wang2022}, where instead of a microwave optical pair $\ket{1_m}\ket{1_o}$, the process generates $\ket{N_m}\ket{N_o}$ pairs, where $N>1$. In the absence of number-resolving detectors, an N-photon generation event may satisfy the one-click condition of the protocol, generating a false positive event. Furthermore, the presence of multi-photon Fock states in the transducer's microwave cavity may lead to multi-photon absorption events in the superconducting qubit, and other deleterious behaviors, such as qubit dephasing and spurious excitations due to residual population of the microwave cavity. Following \cite{Wang2022}, we list the infidelity contributions from multi-photon excitation events below. 

Probability that node A(B) emits a multi-photon state and node B(A) emits zero photons, and the emitted photon reaches the BSA:
\begin{equation}
P_{multi, 0} = P_0(1-P_0-P_1)\eta_l   
\end{equation}
There are two such contributions. There are also two contributions from single photon emission from node A(B), and multi-photon emission from node B(A), where one photon packet is lost and the other reaches the BSA:
\begin{equation}
P_{multi, 1} = P_1(1-P_0-P_1)\eta_l(1-\eta_l)   
\end{equation}
Finally, it may happen that both nodes undergo multi-photon emission, and one of the emitted photon packets is lost before reaching the BSA:
\begin{equation}
P_{multi, multi} = (1-P_0-P_1)^2\eta_l(1-\eta_l).   
\end{equation}
Putting it together, we get the infidelity expression:
\begin{multline}
    P_{in} = P_1^2 \eta_l (1-\eta_l) + \\
    2P_1(1-P_1-P_0)\eta_l + \\
    2P_1 (1-P_0-P_1)\eta_l(1-\eta_l) + \\
    (1-P_0-P_1)^2\eta_l(1-\eta_l).
\end{multline}
It's important to note that multi-photon events impact the \emph{entanglement} fidelity of the transducer, in addition to the readout fidelity of the heralding protocol. This caveat is discussed further below. Going beyond this analysis to include additional contributions to infidelity from thermal noise present in the transducer is a nontrivial task, and is explored in reference \cite{Zhong2020a}.

Putting aside multi-photon generation events, thermal quanta lead to additional false positive detection events, as well as additional problematic phenomena. First, the presence of thermal populations in the transducer's modes decreases the achievable entanglement fidelity within the device: thermal quanta reduce the degree of squeezing in the transducer, thereby reducing the purity of the entangled microwave-optical photon pair \cite{Zhong2020a}. Second, since the superconducting qubit is operating in an absorptive manner, it will absorb microwave photons emitted by the transducer, whether those photons were generated via SPDC or via thermal population of the transducer's acoustic mode. Therefore, regardless of the optical photon click pattern observed, the superconducting qubits within each node may be in a state other than what was supposedly heralded.

To make matters worse, when we are operating the superconducting qubit as an absorber, we must properly shape the microwave photon emitted by the transducer to facilitate absorption \cite{Zhong2024}, adding additional complexity to the protocol. Efficiently capturing or detecting an arbitrarily shaped microwave photon with a superconducting qubit remains an outstanding challenge. Given these considerations, in addition to the order of magnitude lower readout fidelity obtained in previous studies \cite{Krastanov2021, Wang2022}, we focus on the direct transduction protocol in this work.

\section{Spatio-temporal properties of emitted optical modes\label{app:spatiotemporal}}
In this section, we take a focused look at the spatio-temporal properties of the photons emitted by the HBAR transducers, and the impact of fabrication imperfections on the achievable entanglement fidelity. Let us begin by considering the linewidths of the telecom photons that are emitted from the two transducers. An emitted photon's linewidth (in frequency) is determined by the overall linewidth of the transducer’s optical cavity, $\Delta \nu$, which is inversely proportional to its quality factor $Q=\nu/\Delta \nu$. For the resonators studied in this work, the frequency separation between separate devices becomes comparable to the linewidth of the optical resonators themselves. Indeed, the linewidth of the micro-ring resonators ranges from about 3-10 pm, or 400 MHz to 1.2 GHz given a 193 THz frequency. This linewidth directly determines the temporal coherence properties of the emitted photons and the spatio-temporal mode structure of our entangled states.

Specifically, the optical cavity linewidth $\kappa\_o = \kappa\_i + \kappa\_{ext} \approx 150 \;\mathrm{MHz}$ (from our Table 1 parameters) corresponds to a photon coherence time $\tau_{coh} \approx 1/(2\pi\kappa_o) \approx 1 \;\mathrm{ns}$. However, the actual duration of the emitted pulse is determined by the transduction process itself, which we found lasts about $\Delta t \approx 125 \;\mathrm{ns}$, much longer than $\tau_{coh}$. This means our pulses are temporally long and spectrally narrow. Each emitted pulse is temporally extended over about 125 coherence times, but occupies only a small fraction ($1/\Delta t \approx 8 \;\mathrm{MHz}$ out of $150\;\mathrm{MHz}$) of the available cavity bandwidth. The pulse is therefore single-mode in frequency, but temporally multi-mode in that it evolves over many cavity lifetimes. 

For our entanglement heralding protocols, this has a few implications. First, for Type-I heralding, the entangled state is distributed across 125 temporal modes. The entanglement exists in a well-defined frequency mode, but is spread out over $125\;\mathrm{ns}$. Second, for Type-II heralding, we can still have confidence that we can distinguish between Early and Late time-bin photons, since the time-bin separation of $1.125 \;\mathrm{\mu s}$ is much longer than both $\Delta t$ and $\tau_{coh}$. 

Second, an important consideration for entanglement protocols is the spectral distinguishability between photons from different network nodes. Our experimental data show that Si\textsubscript{3}N\textsubscript{4} ring resonators on the same chip exhibit frequency variations of $0.87-13 \;\mathrm{GHz}$, which can be accommodated through several approaches: (1) phase-tracked heralding, or (2) frequency-tunable architectures. 
(1): Following Vittorini et al.\cite{Vittorini2014}, heralding protocols can succeed with spectrally distinguishable photons by tracking the time-dependent phase $\ket{\psi}=\ket{01}+e^{-i\Delta\omega\Delta t}\ket{10}$, provided detector timing resolution satisfies $t_r\ll 2\pi/\Delta \omega$. For SNSPDs with a timing jitter of $10\;\mathrm{ps}$, a frequency difference of up to $100\;\mathrm{GHz}$ can be tolerated. For a total transduction time of $\Delta t=125\;\mathrm{ns}$, the minimum total spectral bandwidth of the emitted photon will be $1/\Delta t = 8 \;\mathrm{MHz}$. Our experimental data show that Si\textsubscript{3}N\textsubscript{4} ring resonators exhibit frequency variations of $0.87-13\;\mathrm{GHz}$. While this prevents perfect spectral matching, entanglement protocols can still succeed through phase-tracked heralding. For frequency-mismatched photons with e.g. $\Delta \omega = 0.87-13 \;\mathrm{GHz}$, the beat length between the mismatched photons varies from $23-340\;\mathrm{mm}$. Phase-tracked heralding protocols can accommodate these variations with sufficient detector timing resolution. 
(2) Double-ring transducers enable frequency tuning capabilities to achieve spectral matching between nodes, circumventing the issues associated with single-ring devices. See Appendix \ref{app:distinguish} for an in-depth discussion of this issue as it relates to FBAR transducers and our proposed protocols. 

Next we turn our attention to the spatial properties of the emitted photons. The pulse bandwidth corresponds to a coherence length of $\sim37\;\mathrm{m}$, which is suitable for laboratory-scale quantum networking demonstrations. Active fiber phase stabilization will allow us to extend the range over which we can distribute entanglement using the photons emitted by these FBAR transducers.

We also include details about the spatial mode properties based on the transducer's photonic circuit design, in case these are also desired. The Si\textsubscript{3}N\textsubscript{4} ($n=2.00$ at 1550 nm) optical ring resonators in the FBAR devices have rectangular waveguides with typical sizes of 800 nm height $\times$ 2.1 $\mathrm{\mu m}$ width buried in an SiO\textsubscript{2} ($n = 1.44$ at 1550 nm) cladding. These devices support well-confined optical modes with a mode field diameter of approximately 1.6 $\mathrm{\mu m}$. Emitted signals are coupled to bus waveguides with a ring-bus gap on the order of $\sim 100\mathrm{'s}\;\mathrm{nm}$ separation, then off-chip using edge couplers and attached high numerical aperture optical fibers. 

Now let us explore the impacts of these parameters on the entanglement heralding protocols. Spatial mode mismatch between two network nodes may reduce interference visibility, impacting the fidelity ceiling of our protocols. We will define the normalized mode overlap as the integral
\begin{equation}
Overlap = \left|\iint E_1^*(x,y) E_2(x,y) \, dx \, dy\right|^2 \approx \left(\frac{2d_1 d_2}{d_1^2 + d_2^2}\right)^2
\end{equation}
with mode field diameters $d1, \; d2$. For our  Si\textsubscript{3}N\textsubscript{4} ring resonator geometry, if we assume typical fabrication tolerances in our photonic waveguides, e.g. $\pm10\;\mathrm{nm}$ width, $\pm5\;\mathrm{nm}$ height, and $\pm50\;\mathrm{nm}$ ring radius, we still anticipate a spatial mode variation of $\pm 0.5\%$. For small variations where $d1 = d2 + \Delta d$, this leads to an overlap of 
\begin{equation}
Overlap \approx 1- (\Delta d/d)^2 \approx >99.999\% 
\end{equation}
for a mode field diameter $d$ with variance $\Delta d$ between nominally identical transducer devices. In the frequency domain, if we assume a $\pm 5\%$ ($\pm 7.5\;\mathrm{MHz}$), this creates some spectral distinguishability. To maintain a $>99\%$ fidelity ceiling, spatial mode mismatches of up to $\pm 2\%$ and spectral mismatches of $\pm5\%$ are both acceptable. Combined with the narrow spectral bandwidth of the emitted photons, our approach should maintain high interference visibility for quantum local area networks.

\section{\label{app:distinguish}Optical photon distinguishability}
One motivation behind generalizing the impedance matching formalism, as is done in the Main Text, is to generalize the design of impedance-matched devices to those including transducers with two optical modes, including double ring resonator devices. This is of paramount importance, because transducers with double ring resonators can have their hybridized optical modes frequency-tuned, e.g. via a DC bias voltage. This capability can be used to tune two transducers to emit photons with a high degree of spectral in-distinguishability. But a question remains: are single-ring FBAR transducers suitable for the proposed entanglement heralding protocols? For a few reasons, detailed below, they are.
 
Given the high-quality fabrication of Si\textsubscript{3}N\textsubscript{4} integrated photonics available, optical micro-ring resonators for transducers, defined either on the same chip or on separate chips, have highly repeatable resonant features. As a specific example, an entanglement heralding experiment could be performed by separately connecting two remote superconducting qubit packages to two adjacent transducers on the same chip.We measured the optical resonances of many separate single-ring transducers on the same chip. See the table below as an example. Optical resonance values for six transducers, sampled arbitrarily throughout a chip with 56 devices, are listed. 

\begin{table}[h]
\centering
\caption{Device Resonances Closest to 1553 nm}
\begin{tabular}{|c|c|}
\hline
\textbf{Device Number} & \textbf{Resonance Closest to 1553 nm} \\
\hline
A & 1553.0794 nm / 193.03099 THz \\
\hline
B & 1553.3661 nm / 192.99537 THz \\
\hline
C & 1553.4106 nm / 192.98984 THz \\
\hline
D & 1553.3930 nm / 192.99202 THz \\
\hline
E & 1553.1840 nm / 193.01799 THz \\
\hline
F & 1553.3731 nm / 192.99450 THz \\
\hline
\end{tabular}
\label{tab:resonance_data}
\end{table}

Of the devices listed here, Devices B and F have a resonant separation of 870 MHz, and Devices C and D have a resonant separation of 2.18 GHz. Devices A and E have a resonant frequency separation of 13 GHz. These frequency difference values fall well below the 100 GHz upper bound set by the jitter of our single photon detectors. 

An additional consideration is the linewidths of the telecom photons that are emitted from the two transducers. The photon’s linewidth is determined by the overall linewidth of the transducer’s optical cavity, $\Delta \nu$, which is inversely proportional to its quality factor $Q=\nu/\Delta \nu$. For the resonators studied in this work, the frequency separation between separate devices becomes comparable to the linewidth of the optical resonators themselves. Indeed, the linewidth of the micro-ring resonators ranges from about 3-10 pm, or 400 MHz to 1.2 GHz given a 193 THz frequency. Two photons emitted from two of these transducers may therefore have non-negligible spectral overlap, rendering them effectively indistinguishable, irrespective of the discussion above.

But what if we can't generate spectrally identical photons from two single-ring devices? Must we restrict ourselves to double-ring devices whose optical output can be frequency-tuned? As shown experimentally by Vittorini and Hucul \emph{et al.} \cite{Vittorini2014}), it is indeed possible to perform heralding gates between spectrally distinguishable flying optical photons. Their specific demonstration showed that if two quantum memories, \emph{A} and \emph{B}, emit photons with frequencies $\omega_A = \omega_B + \Delta \omega$, the remote memories can be projected into a Bell state using a standard Bell State analyzer. The frequency difference between the photons imposes a time-dependent phase on the resulting Bell state, which also depends on the time $\Delta t$ elapsed following Bell state generation:
\begin{equation}
    \ket{\psi}=\ket{01}+e^{-i\Delta\omega\Delta t}\ket{10} 
\end{equation}
The temporal resolution of the photon detectors determines the precision with which the relative phase can be tracked: if the temporal resolution $t_r$  satisfies $t_r \ll 2\pi /\Delta \omega$, the phase of the entangled memory state is well-defined. While it is preferred to have tunable optical resonators to achieve frequency-indistinguishable emission from two separate transducers, if specific conditions are met, heralding gates should succeed even with non-tunable single-ring transducers using a single pump laser.  For superconducting nanowire single photon detectors with timing jitter values of about 10 ps, photons with frequency differences approaching 100 GHz can be accommodated within heralding gates. With proper phase tracking and quantum state tomography, our proposed protocols should be able to accommodate spectrally distinguishable photons within this limit.

In summary, with no mitigation, frequency mismatches exceeding the photon spectral bandwidth of $8\;\mathrm{MHz}$ will drop the quantum interference visibility to zero, dropping the fidelity to $50\%$, no better than classical correlations. With phase-tracked heralding, the fidelity ceiling for indistinguishable photons is preserved if the timing resolution is sufficient. As mentioned above, for an SNSPD timing jitter of $10\;\mathrm{ps}$, we have, at worst, a resolution requirement of $2\pi/\Delta \omega = 77\;\mathrm{ps}$. This timing resolution is easily achievable for standard SNSPDs. Due to the experimental overhead associated with phase tracking, the timing jitter associated with the distinguishable photon pair may degrade the fidelity ceiling to $90\%$. Finally, with double-ring frequency tuning, spectral in-distinguishability should be maintained, so the anticipated fidelity ceiling should match the original baseline values of $90-98\%$.

\section{\label{app:sensitivity}Sensitivity analysis of matching circuit errors}
For realistic experiments, transducer devices, including impedance-matching superconducting circuits, will be subject to standard fabrication variances. Here, we analyze the impacts of fabrication variances of up to $\pm20\%$ on the matching capacitor $C_T$ and matching inductor $L$, and calculated the impact of these variances on the transducer's figures of merit, as well as the performance of the entanglement heralding protocol. Table \ref{tab:sensitivity_analysis} shows the variance in these figures of merit for worst-case $\pm20\%$ variations in the matching network. Variances in transducer conversion efficiency are about $\sim5\%$ for efficiency-maximized devices and $\sim20\%$ for noise-minimized devices. Added thermal noise values are barely affected ($<1\%$ change), and the conversion bandwidth decreases by about $3\%$ for efficiency-maximized devices and by about $16\%$ for noise-minimized devices. 

Fortunately, the fidelity and entanglement rate of the Type-I and Type-II protocols is barely affected ($<1\%$ change), even with noise-minimized transducers. A summary of the sensitivity analysis as compared to baseline values is included in Table \ref{tab:baseline_sensitivity}. 

\begin{table*}[h]
\centering
\caption{Sensitivity Analysis: ±20\% Variations in Matching Components}
\begin{tabular}{|l|c|c|c|c|}
\hline
\textbf{Parameter} & \multicolumn{2}{c|}{\textbf{Single Ring}} & \multicolumn{2}{c|}{\textbf{Double Ring}} \\
\cline{2-5}
& \textbf{Max Eff} & \textbf{Min Noise} & \textbf{Max Eff} & \textbf{Min Noise} \\
\hline
\hline
\multicolumn{5}{|c|}{\textbf{Transducer Performance}} \\
\hline
Conversion Efficiency ($\Delta\eta$, \%) & $-5.27$ & $-20.00$ & $-5.27$ & $-20.00$ \\
\hline
Thermal Noise ($\Delta n_{th}$, \%) & $-0.03$ & $-0.61$ & $-0.03$ & $-0.61$ \\
\hline
Bandwidth ($\Delta BW$, \%) & $-2.64$ & $-15.66$ & $-2.64$ & $-15.66$ \\
\hline
\hline
\multicolumn{5}{|c|}{\textbf{Type-I Protocol}} \\
\hline
Fidelity ($\Delta F_{I}$, \%) & $-0.06$ & $-0.14$ & $-0.06$ & $-0.14$ \\
\hline
Rate ($\Delta R_{I}$, \%) & $-0.59$ & $-0.51$ & $-0.59$ & $-0.51$ \\
\hline
\hline
\multicolumn{5}{|c|}{\textbf{Type-II Protocol}} \\
\hline
Fidelity ($\Delta F_{II}$, \%) & $-0.59$ & $-0.35$ & $-0.59$ & $-0.35$ \\
\hline
Rate ($\Delta R_{II}$, \%) & $-0.59$ & $-0.51$ & $-0.59$ & $-0.51$ \\
\hline
\end{tabular}
\label{tab:sensitivity_analysis}
\end{table*}

\begin{table*}[h]
\centering
\caption{Baseline Values and Sensitivity Analysis Summary}
\begin{tabular}{|l|c|c|c|c|}
\hline
\textbf{Parameter} & \multicolumn{2}{c|}{\textbf{Single Ring}} & \multicolumn{2}{c|}{\textbf{Double Ring}} \\
\cline{2-5}
& \textbf{Max Eff} & \textbf{Min Noise} & \textbf{Max Eff} & \textbf{Min Noise} \\
\hline
\hline
\multicolumn{5}{|c|}{\textbf{Baseline Values}} \\
\hline
Conversion Efficiency (\%) & $42.20$ & $28.69$ & $42.16$ & $28.66$ \\
\hline
Thermal Noise (quanta) & $2.20 \times 10^{-2}$ & $6.11 \times 10^{-3}$ & $2.21 \times 10^{-2}$ & $6.11 \times 10^{-3}$ \\
\hline
Type-I Fidelity (\%) & $90.68$ & $86.52$ & $90.64$ & $86.45$ \\
\hline
Type-I Rate (kHz) & $168.9$ & $161.8$ & $169.3$ & $162.3$ \\
\hline
Type-II Fidelity (\%) & $92.82$ & $97.90$ & $92.82$ & $97.90$ \\
\hline
Type-II Rate (kHz) & $37.99$ & $36.40$ & $38.10$ & $36.52$ \\
\hline
\hline
\multicolumn{5}{|c|}{\textbf{Robustness Assessment}} \\
\hline
Efficiency Sensitivity & Moderate & High & Moderate & High \\
\hline
Noise Sensitivity & Low & Moderate & Low & Moderate \\
\hline
Protocol Fidelity Impact & Minimal & Minimal & Minimal & Minimal \\
\hline
Overall Robustness & \textbf{Good} & \textbf{Fair} & \textbf{Good} & \textbf{Fair} \\
\hline
\end{tabular}
\label{tab:baseline_sensitivity}
\end{table*}

\bibliography{library}

\end{document}